\documentclass{article}
\usepackage{amsthm}
\usepackage{graphicx}
\usepackage{dcolumn}
\usepackage{bm}
\usepackage{float}
\usepackage[margin=1.2in]{geometry}
\usepackage{comment}
\usepackage{amsmath}
\usepackage{xcolor}
\usepackage{hyperref}
\usepackage{cite}
\begin{document}
\title{Quasi-symmetries in complex networks: a dynamical model approach}

\author{Gemma Rosell-Tarrag\'{o}\footnote{Corresponding author. Email:gemmaroselltarrago@gmail.com} $^{1,2}$ \and Albert D\'{i}az-Guilera$^{1,2}$}
\date{%
    \small{$^1$Departament de F\'{i}sica de la Mat\`{e}ria Condensada, Universitat de Barcelona, Mart\'{i} i Franqu\`{e}s 1, 08028 Barcelona, Catalonia, Spain\\%
    $^2$Universitat de Barcelona Institute of Complex Systems (UBICS), Universitat de Barcelona, Barcelona, Catalonia, Spain\\[2ex]%
}}
\maketitle

\begin{abstract}
The existence of symmetries in complex networks has a significant effect on network dynamic behaviour. Nevertheless, beyond topological symmetry, one should consider the fact that real-world networks are exposed to fluctuations or errors, as well as mistaken insertions or removals. Therefore, the resulting approximate symmetries remain hidden to standard symmetry analysis - fully accomplished by discrete algebra software. There have been a number of attempts to deal with approximate symmetries. In the present work we provide an alternative notion of these weaker symmetries, which we call `quasi-symmetries'. Differently from other definitions, quasi-symmetries remain free to impose any invariance of a particular network property and they are obtained from the phase differences at the steady-state configuration of an oscillatory dynamical model: the Kuramoto-Sakaguchi model. The analysis of quasi-symmetries unveils otherwise hidden real-world networks attributes. On the one hand, we provide a benchmark to determine whether a network has a more complex pattern than that of a random network with regard to quasi-symmetries, namely, if it is structured into separate quasi-symmetric groups of nodes. On the other hand, we define the `dual-network', a weighted network (and its corresponding binnarized counterpart) that effectively encodes all the information of quasi-symmetries in the original network. The latter is a powerful instrument for obtaining worthwhile insights about node centrality (obtaining the nodes that are unique from that act as imitators with respect to the others) and community detection ( quasi-symmetric groups of nodes).
\end{abstract}

\section{Introduction}
\label{sec:Introduction}
Complex networks - from biological networks such as the brain connectome or regulatory networks to social and technological networks, like scientific collaboration networks or the Internet \cite{Caldarelli2007,Barabasi2009,Caldarelli2007a,PastorSatorras2004, Newman2001} - are widely used to model the structure and behaviour of complex systems. Despite these apparently diverse networks are unique in its nature, many studies have shown that they share a number of properties, which distinguish them from other mathematical graphs of interest. Such common features include the heterogeneity in its node degrees, captured by a power-law distribution, high clustering coefficients, and the `small-world' property, among others \cite{Watts1998b,Albert2002,Boccaletti2006,Strogatz2001,Girvan}. Additionally, a certain degree of symmetry is also an attribute of real-world networks \cite{Smith2019,Sanchez-Garcia2020}. The study of the symmetries of a network is of great relevance for several reasons: it may help us to have a better understanding of the formation of certain real-world networks, they can also provide information about node function, and have an effect on network redundancy and robustness. Moreover, symmetries are known to influenciate the outcome of network dynamics, such as synchronization or controllability \cite{Liu2011,Liu2013,Whalen2015,Nicosia2013a,Pecora2014,Jiang2016}.

The notion of `symmetry' or `invariance' includes several specifications depending on the field it is applied \cite{Zee1986-ZEEFST}. Mathematically, a symmetric transformation, or a symmetry is the set of transformations that leaves an object invariant or unmodified \cite{Lockwood1978}. Differently than continuous transformations, such as a translation or a rotation applied to a geometric shape, symmetries in complex networks are necessarily discrete transformations applied to graphs, which are defined as discrete entities. Importantly, graphs are topological objects and generally, their properties are independent of the positions of vertices or lenghts of the links. For this reason, a geometric transformation of their components has no effect on the topology, but to the visualization of the graph . In a different way, a topological transformation of a graph maps each vertex to another one as a permutation. Finally, the set of permutations of a graph that leaves the topology invariant are the automorphisms of the graph (in Section \ref{sec:Symmetries} the notion of symmetries in complex networks is explained in depth). Other types of symmetries that may be present in graphs are scale invariance or translational symmetries, which are not considered in the present work \cite{Garlaschelli2010}.

Built on the standard notion of graph symmetry that we have reported, i.e, topological or structural symmetry, other weaker or approximate symmetries may be present in real-world networks. Despite they are not included in the finite number of automorphisms of graphs, they indeed play an important part in determining the network behaviour \cite{Garlaschelli2010,Stewart2004,Olver2016}. Alternatives for approximate symmetries in graphs include `near' symmetries and `stochastic symmetries' \cite{Smith2019, Holme2006}. A `near' symmetry is described in terms of properties of the network that are left unchanged when some other transformation is applied on the network. Examples include whether two nodes have the same degree, and/or the same number of second neighbours, and/or the same local clustering coefficient. A more relaxed condition consists in whether two nodes are `statistically' equivalent, that is, whether these topological properties are the same in an average sense. The permutation of statistically equivalent nodes are called stochastic symmetries and they result in a family of statistically equivalent networks with the same statistical properties \cite{Garlaschelli2010}.

The given alternatives to perfect or standard topological symmetries in graphs are of great interest as small fluctuations or errors may be present when constructing the graphs, as well as additional and/or missing links could be included/removed. The resulting graphs or networks may lead to very significant changes in the analysis of topological symmetries, as many of them will remain hidden due to its approximate nature.

In the present work, and in line with the analysis of approximate symmetries, we propose a different extension of the latter, which we call `quasi-symmetries'. This alternative definition of weaker symmetries remains free to impose any invariance of a particular topological property. Quasi-symmetries are obtained from the network as an extension to structural equivalence: structural or topological similarity is derived for all pair of nodes from an oscillatory dynamical model: the Kuramoto-Sakaguchi model \cite{Sakaguchi1986}. According to this model, all nodes are considered as individual phase-oscillators that are coupled with its neighbours by a sinus function of its phase difference. The phase differences between them at the steady-state configuration determine the degree of structural similarity, as shown in Section \ref{subsec:SymmetriesDetection}. The analysis of quasi-symmetries provides insights to otherwise hidden properties of real-world networks. Firstly, we explore the distributions of structural similarity among all pairs of nodes and we find a benchmark to determine whether a network has a more complex pattern than that of a random network concerning quasi-symmetries . Secondly, we define the `dual network', a weighted network (and its corresponding binnarized counterpart) that effectively encodes all the information of quasi-symmetries in the original one. The dual network allows for the analysis of centrality measures and community detection. The first informs us about the nodes that play a unique role in the network or those that behave similarly to many other nodes. The latter results to a classification of nodes into quasi-symmetric communities, the natural extension of the automorphism group orbits (structurally symmetric nodes) of a network.

The paper is organized as follows: section \ref{sec:Symmetries} provides a short review of the notion of symmetries in complex networks, focusing on the concept of the orbits of a network. In section \ref{subsec:BuildingSymmetries} we explain a methodology to generate synthetic networks with controlled symmetries, based on Ref.\cite{Klickstein2018}. An alternative methodology to detect structural or perfect symmetries is explained in section \ref{subsec:SymmetriesDetection}. Section \ref{sec:Quasi-symmetries} is the central part of the paper and also our main contribution to the literature. Quasi-symmetries are explained in detail through its construction (in section \ref{subsec:BuildingQuasi-symmetries}), characterization (in section \ref{subsec:characterization}) and definition of the dual network (in section \ref{sec:dualNetwork}). Further mathematical derivations and large visualizations of real-world networks can be found in the Appendix.

\section{Symmetries in complex networks}
\label{sec:Symmetries}
A network or, mathematically, a simple graph, $\mathcal{G}(\mathcal{V},\mathcal{E})$, consists of a set of nodes, $\mathcal{V}(\mathcal{G})$, linked by a set of edges $\mathcal{E}(\mathcal{G})$. A network of $n$ nodes, labelled from $0$ to $n-1$, can be represented by its adjacency matrix, $A$, a $n \times n$ matrix with $a_{ij}=1$ if there is a link between nodes $i$ and $j$ and $a_{ij}=0$ otherwise. A permutation, or relabelling, of the nodes of a network can be written as $\pi(\mathcal{V}):\{0,1,...,n-1\}\to \{\pi(0),\pi(1),...,\pi(n-1)\}$ where, for instance, node $0$ changes to $\pi(0)$. Equivalently, a permutation can be represented in a two-line form as follows, 
\begin{equation}\pi(\mathcal{V}):
\begin{pmatrix}
0 & 1 & ... & n-1 \\
\pi(0) & \pi(1) & ... & \pi(n-1)
\end{pmatrix}
\label{eq:permutation}    
\end{equation}
$P_{\pi}$ is a square matrix  that corresponds to the permutation $\pi(\mathcal{V})$ and is obtained by permuting the columns of the identity matrix, i.e., the element $p_{ij}=1$ if $\pi(i)=j$ and $0$ otherwise.

The concept of network symmetry is akin to the mathematical definition of a graph automorphism, which is a permutation of the network nodes but preserving adjacency. In other words, neighbouring nodes still remain neighbours after the permutation is applied. Namely, a graph automorphism $\sigma(\mathcal{V})$ is a permutation of the vertices $\sigma(\mathcal{V})$ such that $(\sigma(i),\sigma(j))$ is an edge only if $(i,j)$ is an edge: the set of edges is preserved. Consequently, the permutation matrix corresponding to a graph automorphism or a symmetry, $P_{\sigma}$, commutes with the adjacency matrix of the network.
\begin{equation}
AP_{\sigma}=P_{\sigma}A
\label{eq:commute}
\end{equation}
The set of all the symmetries of a graph form the automorphism group of the graph, $Aut(\mathcal{G})$. In Reference \cite{Erdos1963}, a graph is defined as symmetric when there exits at least a non-identical permutation of its vertices that leaves the graph invariant or, equivalently, the group of its automorphisms has a degree greater than 1.

\begin{figure}[H]
    \centering
    \includegraphics[width=0.6\textwidth]{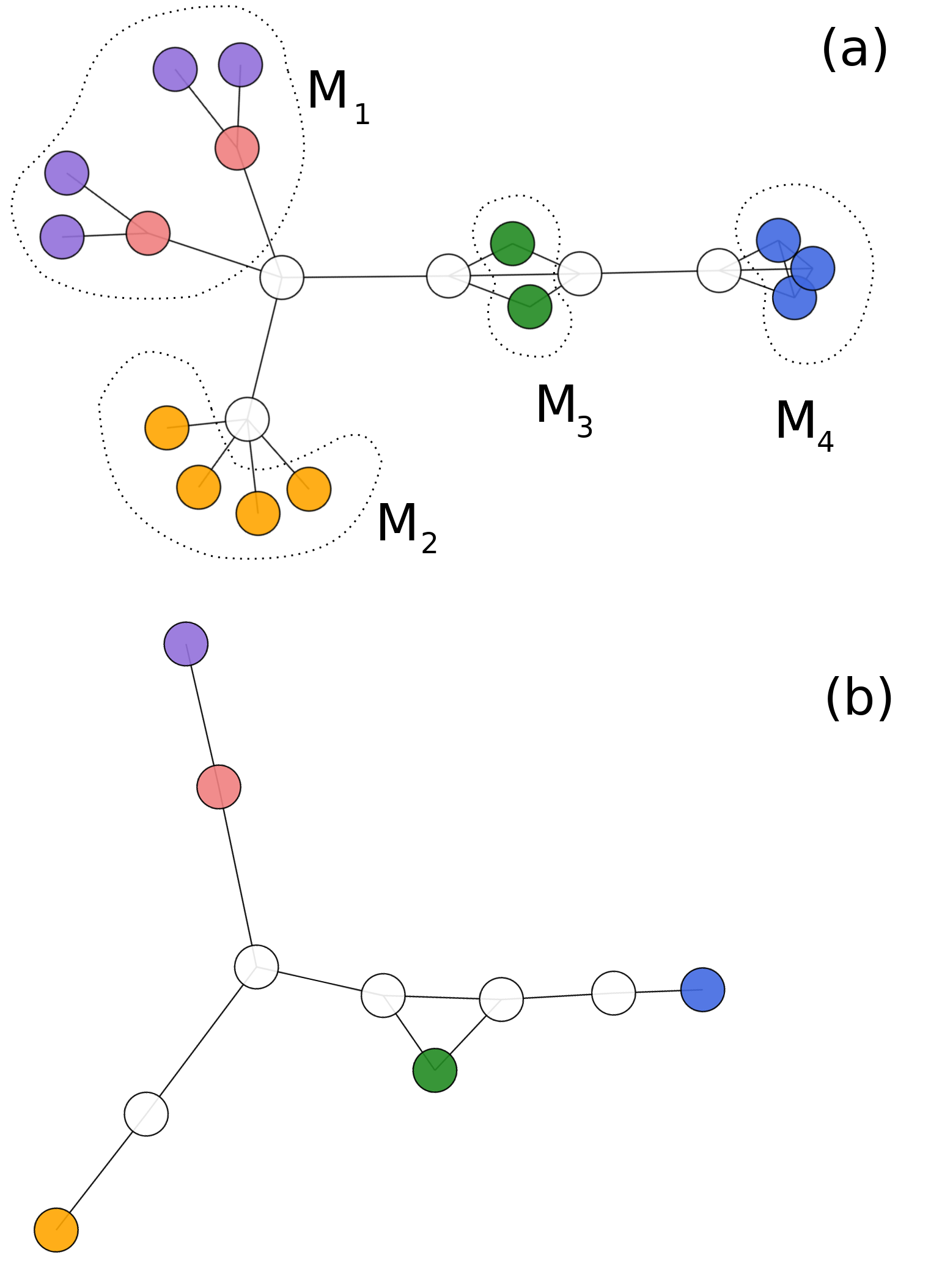}
    \caption{(Color online) Geometric decomposition into the asymmetric core and four symmetric motifs of a network, in panels (b) and (a), respectively. Nodes are colored by orbit and fixed points are in white color. Motifs $M_2$, $M_3$ and $M_4$ correspond to BSMs and motif $M_1$ is non-basic.}
\label{fig:geometricDecomposition}
\end{figure}

The set of vertices can be split into the core of fixed points, $V_0$, that is, vertices which are moved by none of the automorphisms of $Aut(\mathcal{G})$, and the vertex set of symmetric motifs, $M_i$. This partition is called the geometric decomposition of the network and can be written as
\begin{equation}
V = V_0 \cup M_1\cup ...  \cup M_m
\end{equation}
being $m$ the number of symmetric motifs. Each symmetric motif can be further partitioned into clusters. Two nodes, $v_i$ and $v_j$, belong to the same cluster if $\sigma(v_i)=v_j$ and conversely, where $\sigma \in Aut(\mathcal{G})$. Clusters are alternatively called orbits induced by $Aut(\mathcal{G})$. The vertices or nodes of the same orbit are structurally indistinguishable and play the same structural role in the network (nodes are colored by orbit in Fig. \ref{fig:geometricDecomposition}).
We can classify symmetric motifs into two types: basic and complex. Basic symmetric motifs (BSMs) are made of one or more orbits of the same number of vertices (motifs $M_2$, $M_3$ and $M_4$ in Fig. \ref{fig:geometricDecomposition}) and complex symmetric motifs are hardly found in real-world networks, and they are typically branched trees (motif $M_1$ in Fig. \ref{fig:geometricDecomposition})\cite{MacArthur2008,Ben2011,Sanchez-Garcia2020}.
The detection of graph automorphisms and the corresponding geometric decomposition of a network is vastly used to simplify the topology of the network by compressing redundant information. Moreover, the basic structural properties of the network can be derived only from the geometric decomposition of the graph or the so called quotient graph. Network eigenvalues are an example of it.

In the present work we are interested in detecting the nodes that are structurally equivalent, that is, nodes that play the same role in a network and therefore, we will be detecting the orbits generated by the automorphism group of a network, $Aut(\mathcal{G})$. Notice that a symmetric motif may be subdivided into several orbits and that the isolated permutation of two nodes belonging to the same orbit needs not correspond to an automorphism of the network.

The notion of `structural equivalence' or a pair of nodes being structurally equivalent is alternatively defined in the social sciences as: if two nodes have exactly the same set of neighbours, regardless of whether they are neighbours of each other, then a permutation between them exists such that the network remains unchanged. Notice, however, that this definition is more restrictive that two (or more) nodes being structurally equivalent as long as they belong to the same orbit, which may not share the same neighbours, however.
\subsection{Generation of symmetric networks}
\label{subsec:BuildingSymmetries}
By examining the automorphism group of real-world networks, several studies show that real networks, unlike random graphs, contain a large amount of symmetries, namely, network motifs\cite{Alon2007}. This is partly due to the fact that symmetry can arise from growth processes present in nature. However, the availability of real network datasets is often scarce, especially, when looking for enough variability regarding symmetry. Alternately, we can use random graphs generating models, such as Erd\"{o}s-R\'{e}nyi, Watts-Strogatz or Barab\'{a}si-Albert, but these models do not generate graphs with symmetries, and hence we should turn to regular graphs in order to work with symmetries. Such motifs are however trivial and easy-to-identify by visual inspection.

In the present work we will use an algorithm that is able to generate graphs with any desired symmetry pattern \cite{Klickstein2018}. Hereafter, we provide a schematic explanation of the algorithm and the main required concepts.

An \textit{equitable partition}(EP) of the nodes divides the graph into non-overlapping clusters of nodes, $\{C_i\}$, such that the number of connections to $C_j$ from any node $v \in C_i$ only depends on $i$ and $j$, that is, their corresponding clusters \cite{Schaub2016}. 

The automorphism group, $Aut(\mathcal{G})$, of a graph induces an equitable partition of nodes, where the clusters of the EP are the orbits generated by $Aut(\mathcal{G})$. 

\begin{figure}[H]
    \centering
    \includegraphics[width=0.6\textwidth]{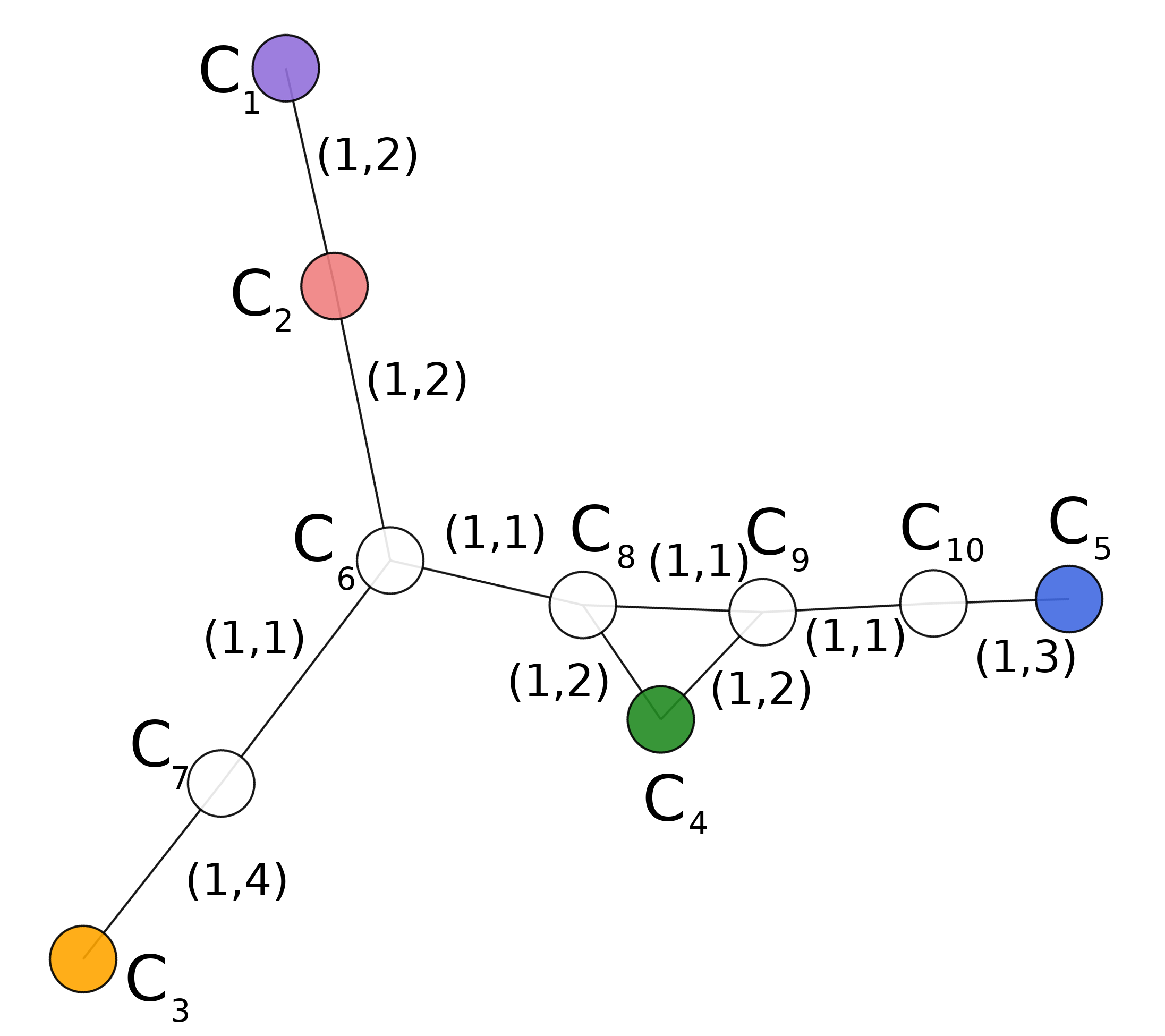}
    \caption{(Color online) Quotient graph, $\mathcal{Q}$, corresponding to the network in Fig. \ref{fig:geometricDecomposition}(a). $C_i$ correspond to the different clusters or quotient nodes of $\mathcal{Q}$. Using the components described in Eq.(\ref{eq:Q}), $\vec{n}=(4,2,4,2,3,1,1,1,1,1)$, $\vec{s}=(0,0,0,0,2,0,0,0,0,0)$ and the weights of the quotient edges are included in the figure using the notation $(f_{jk},f_{kj})$ with $j<k$.\label{fig:structureQ}}
\end{figure}

An equitable partition of a graph can be represented by its \textit{quotient graph}, $\mathcal{Q}$. The quotient graph of an EP consists of five components:
\begin{equation}
\label{eq:Q}
    \mathcal{Q}=\{\mathcal{C}, \mathcal{F}, \vec{n}, \vec{s}, \vec{f}\}
\end{equation}
$\mathcal{Q}$ is made of $p$ quotient nodes and $q$ quotient edges. $\mathcal{C}$ represents the set of clusters or quotient nodes and $\mathcal{F}$ represents the set of quotient edges that link the clusters of the EP. The integer vector $\vec{n}$ of length $p$ contains the size of each cluster or quotient nodes, while the integer vector $\vec{s}$ of length $p$ represents the intra-cluster degree of each cluster, that is, the number of edges of a node with all the others within the same cluster (which is a shared number for all nodes in the cluster). The integer vector $\vec{f}$ of length $2q$ consists of pairs of quotient edge weights assigned to each quotient edge $(C_i,C_j)\in\mathcal{F}$ as $(f_{jk},f_{kj})$ defined as
\begin{eqnarray}
\label{eq:edgesWeights}
    f_{jk}=\sum_{v_a\in C_k}A_{ia}, \ v_i\in C_j \nonumber\\
    f_{kj}=\sum_{v_a\in C_j}A_{ia}, \ v_i\in C_k
\end{eqnarray}
In Fig. \ref{fig:structureQ} we show the quotient graph corresponding to the network in Fig. \ref{fig:geometricDecomposition}(a).
However, not all quotient graphs are feasible, that is, not all combination of the components of $\mathcal{Q}$ described in Eq.(\ref{eq:Q}) represents some original graph. The authors of the algorithm for generating symmetric graphs \cite{Klickstein2018} take into account several constraints that must be considered: firstly and in order to satisfy that the number of nodes of each cluster, $n_i$, can satisfy the connectivity requirements implied by $s_i$, the following restrictions have to be met:
\begin{equation}
\label{eq:constrain1}
    \mod (n_i s_i,2)=0, \ n_i \ge s_i+1
\end{equation}
In addition, the number of edges, $m_{jk}$, going through two linked clusters, $C_j$ and $C_k$, must be consistent:
\begin{equation}
\label{eq:constrain2}
    n_j f_{jk} = n_kf_{kj} = m_{jk}
\end{equation}
which also imply that there must be enough nodes in cluster $C_k$ to satisfy the demands of each node in cluster $C_j$ and the other way round:
\begin{equation}
    \label{eq:constrain3}
    n_k\ge f_{jk}, \ n_j \ge f_{kj}
\end{equation}
The constraints defined in Eqs.(\ref{eq:constrain1})-(\ref{eq:constrain3}) gathers the conditions that a quotient graph, $\mathcal{Q}$, must meet to be feasible. In addition, one could construct a representative graph $\mathcal{G}$ from a given quotient graph. This last implication is of particular relevance as the authors suggest a methodology to obtain samples of symmetric graphs that fulfill the requirements of a particular quotient graph. We will briefly present the steps of the algorithm, but we encourage the reader to find all the details in the cited work\cite{Klickstein2018}.

The required input consists of the sets $\vec{n}$, $\vec{s}$ and $\mathcal{F}$, together with the number of quotient nodes and quotient edges, $p$ and $q$, respectively. The resulting graph, $\mathcal{G}$, has $\sum_{i=1}^pn_i$ nodes and $\sum_{i=1}^p\frac{n_is_i}{s}+\sum_{(C_j,C_k)}\in \mathcal{F}n_if_{jk}$ edges. They next propose a method to provide a proper choice of the quotient edges weights, without having first made sure that the constraints defined in Eqs.(\ref{eq:constrain1})-(\ref{eq:constrain3}) are met. They divide the set of $\mathcal{G}$ edges into the intra-cluster and the inter-cluster edge sets and suggest a wiring scheme for the edges, based on mathematical proofs. The equitable partition induced by the created $Aut(\mathcal{G})$ is verified by using software  \texttt{Nauty}\cite{McKay2014}.
\subsection{Detection of symmetries: a dynamic model approach}
\label{subsec:SymmetriesDetection}
There are many discrete algebra software that is able to determine the automorphism group, that is, the symmetries, of a graph as well as to extract the orbits that locate the nodes in each cluster. \texttt{Saucy3}\cite{saucy}, \texttt{GAP}\cite{gap} or \texttt{Nauty}\cite{McKay2014} are some examples. We are however interested in constructing a framework that enables the detection of, not only perfect symmetries, but what we will call \textit{quasi-symmetries} (See Section \ref{sec:Quasi-symmetries}). 

To this end, we present an alternative method to detect the orbits of a network by using the steady state of a dynamic model: the Kuramoto-Sakaguchi model with homogeneous phase lag. 
Consider the dynamics of $N$ identical phase oscillators $\theta_i$, for $i=\{1,...,N\}$, coupled in a network whose evolution is governed by
\begin{equation}
\label{eq:KSmodel}
    \dot{\theta}_i=\omega+K\sum_j A_{ij}\sin(\theta_j-\theta_i-\alpha), \ j\in \Gamma_{i}
\end{equation}
Eq.(\ref{eq:KSmodel}) corresponds to the Kuramoto-Sakaguchi model (1986)  \cite{Sakaguchi1986}, which adds to the original Kuramoto model (1975)  \cite{Kuramoto1975,Acebron2005,Arenas2008} a homogeneous phase lag, $\alpha$, between nodes that promotes a phase shift between oscillators. Each unit is influenced directly by the set of its nearest neighbours via the adjacency matrix of the network corresponding to the system, $\mathcal{G}(\mathcal{V}, \mathcal{E})$. The coupling strength, $K>0$, adjusts the intensity of such interactions, $\Gamma_i$ is the set of neighbours of node $i$ and $\omega$ is the natural frequency of each unit, which we consider to be homogeneous among oscillators. 

It has been shown that, as long as $|\alpha| < \pi/2$, the system is not chaotic and it becomes synchronized to a resulting frequency \cite{Sakaguchi1986}. In the dynamics described in Eq.(\ref{eq:KSmodel}), the frustration parameter, $\alpha$, forces the system to break the otherwise original fully synchronized state, that is, phase synchronization. However, partial synchronization is conserved for nodes belonging to the same orbit in the network \cite{Nicosia2013a,Nishikawa2016}. We hereafter provide a proof of this last statement. Let us first derive the analytical solution of the phases in the steady state.

If the system reaches the synchronized state and $\alpha$ is small enough, Eq.(\ref{eq:KSmodel}) can be linearized and the values of the phases at any time in the steady-state are given by
\begin{equation}
    \sum_jL_{ij}\theta_j = \alpha (\left\langle d \right\rangle-d_i)
\end{equation}where $d_i$ is the degree of the $i$th node and the Laplacian matrix of the network $\mathcal{G}$ is defined as
\begin{equation}
    \label{eq:laplacianMatrix}
    L \equiv D - A
\end{equation}where $A$ is the adjacency matrix of the network and $D$ is the diagonal matrix $[D]_{ij} = d_{i}\delta_{ij}$ and $d_i$ is the degree of the $i$th node. Equivalently, $L_{ij}=d_{i}\delta_{ij}-A_{ij}$. In matrix notation,
\begin{equation}
\label{eq:phasesMatrixNotation}
    L \vec{\theta} = \alpha (\left\langle d \right\rangle \vec{1}_n-\vec{d})
\end{equation}where $\vec{[d]}_i = d_i$ (See a detailed proof in the Appendix section). In a connected network, $L$ has one null eigenvalue. Consequently, Eq.(\ref{eq:phasesMatrixNotation}) is singular. Nonetheless, we can solve it by computing the phase difference between each node and a node which we choose as reference. Hence, 
\begin{equation}
    \label{eq:referenceNode}
    \phi_i \equiv \theta_i-\theta_R
\end{equation}where $R$ is the index of the reference node and its corresponding $\theta_R$ is left as a free variable. Obviously, $\theta_R=0$. The new system can be written as
\begin{equation}
\label{eq:phasesReduced}
    \tilde{L} \vec{\phi} = \alpha (\left\langle d \right\rangle \vec{1}_{n-1}-\tilde{\vec{d}})
\end{equation}where $\tilde{L}$, the so called reduced Laplacian \cite{Nicosia2013a,Rosell-Tarrago2020}, is obtained by removing the $R$th row and column of $L$, although the result does not depend on which row we remove. Similarly, the vector $\tilde{\vec{d}}$ is obtained by removing the $R$th element of $\vec{d}$. Finally, the phases with respect to a reference node in the frequency synchronized steady state of the Kuramoto-Sakaguchi model are given by
\begin{equation}
\label{eq:phasesSolution}
    \vec{\phi} = \alpha \tilde{L}^{-1} (\left\langle d \right\rangle \vec{1}_{n-1}-\tilde{\vec{d}})
\end{equation}
We next show that the phases of nodes belonging to the same orbit will be equal at any time.

If $P$ corresponds to the permutation matrix of an automorphism $\sigma \in Aut(\mathcal{G})$, then Eq.(\ref{eq:commute}) is true. The Laplacian matrix of the network, $L$, also commutes with the permutation matrix, as
\begin{equation*}
    PL = P(D-A) = PD-PA
\end{equation*}
We already know that $P=P_{\sigma}$ commutes with $A$, as $\sigma\in Aut(\mathcal{G})$. $P$ also commutes with $D$ on account of the general statement that any diagonal matrix with equal values for all elements corresponding to the same orbit of the automorphism permutes with the corresponding permutation matrix (See the Appendix section for a detailed proof and \cite{Sanchez-Garcia2020} for a generalization of this result). All nodes belonging to the same orbit have the same degree, and hence, $D$ meets the required conditions so as to permute with $P$. Hence,
\begin{equation}
\label{eq:laplacianCommutes}
    PL = PD-PA = DP-AP=(D-A)P=LP
\end{equation}
If we left-multiply Eq.(\ref{eq:phasesMatrixNotation}) by $P$ we get
\begin{equation*}
    PL \vec{\theta} = \alpha (\left\langle d \right\rangle P\vec{1}_n-P\vec{d}) = \alpha (\left\langle d \right\rangle \vec{1}_n-\vec{d})
\end{equation*}as symmetric nodes have the same degree ($P\vec{d}=\vec{d}$). In addition, $PL=LP$, as derived in Eq.(\ref{eq:laplacianCommutes}). Consequently,
\begin{equation}
\label{eq:permutePhases}
    LP \vec{\theta} = L\vec{\theta}
\end{equation}
Similarly as done in Eq.(\ref{eq:phasesReduced}), we define $\tilde{P}$ as $P$ with the removal of the $R$th row and column and Eq.(\ref{eq:permutePhases}) turns to
\begin{equation}
\label{eq:permutePhasesReduced}
    \tilde{L}\tilde{P} \vec{\phi} = \tilde{L}\vec{\phi}
\end{equation}
Now, the inverse of $\tilde{L}$ exists and we can left-multiply Eq.(\ref{eq:permutePhasesReduced}) by $\tilde{L}^{-1}$, leading to
\begin{equation}
\label{eq:proofPhasesSymmetric}
    \tilde{P} \vec{\phi} = \vec{\phi}
\end{equation}
Since $\tilde{P} \vec{\phi}$ corresponds to the permutation of the phases of symmetric nodes, Eq.(\ref{eq:proofPhasesSymmetric}) implies that the phases of nodes belonging to the same orbit (those permuted within an automorphism) are equal at any time.

 The reverse conditional statement is always true with the exception of a very unlikely case. Only when two nodes $i$ and $j$ that have different degrees, i.e., $d_i \neq d_j$, verify this very restrictive condition (see Appendix \ref{ap:biconditional})
\begin{equation}
\frac{\sum_k [\tilde{L}^{-1}]_{ik}}{\sum_k [\tilde{L}^{-1}]_{jk}} = \frac{\left\langle d \right\rangle - d_i}{\left\langle d \right\rangle - d_j}
\label{eq:bi-conditional_1}
\end{equation}
and, additionally the degrees of both nodes meet the inequality 
\begin{equation}
d_i \geq \left\langle d \right\rangle \text{ and } d_j \geq \left\langle d \right\rangle \text{ or } 0 < d_i \leq \left\langle d \right\rangle \text{ and } 0 < d_j \leq \left\langle d \right\rangle 
\label{eq:bi-conditional_2}
\end{equation}
then the two considered nodes can have the same phases despite not belonging to the same orbit.

Nevertheless, we note that the condition expressed in Eq.(\ref{eq:bi-conditional_1}) represents a highly unlikely event and hence would require a very fine tuning of the degree sequence of the corresponding (weighted) network. Moreover, from a probabilistic perspective, the probability that a continuos random variable takes a specific value is zero and so is the chance that the quotient of weighted degrees in Eq.(\ref{eq:bi-conditional_1}), resulting from a non-linear transformation, takes a particular value. Henceforth we will assume that the bi-conditional stated as `Nodes that have the same phases $ \Longleftrightarrow$  Nodes that belong to the same orbit' is effectively true.

\color{black}In this section we have proved that the phases at the steady state of the Kuramoto-Sakaguchi model with homogeneous natural frequencies and phase lag parameters capture the clusters of nodes corresponding to the orbits of the network. Therefore, a straightforward method to detect the orbits of a network is computing the phases analytically as in Eq.(\ref{eq:phasesSolution}) and classify nodes into clusters according to their values. Nodes with equal values of $\phi$ belong to the same orbit. 

\begin{figure}[H]
\centering
\includegraphics[width = 0.7\textwidth]{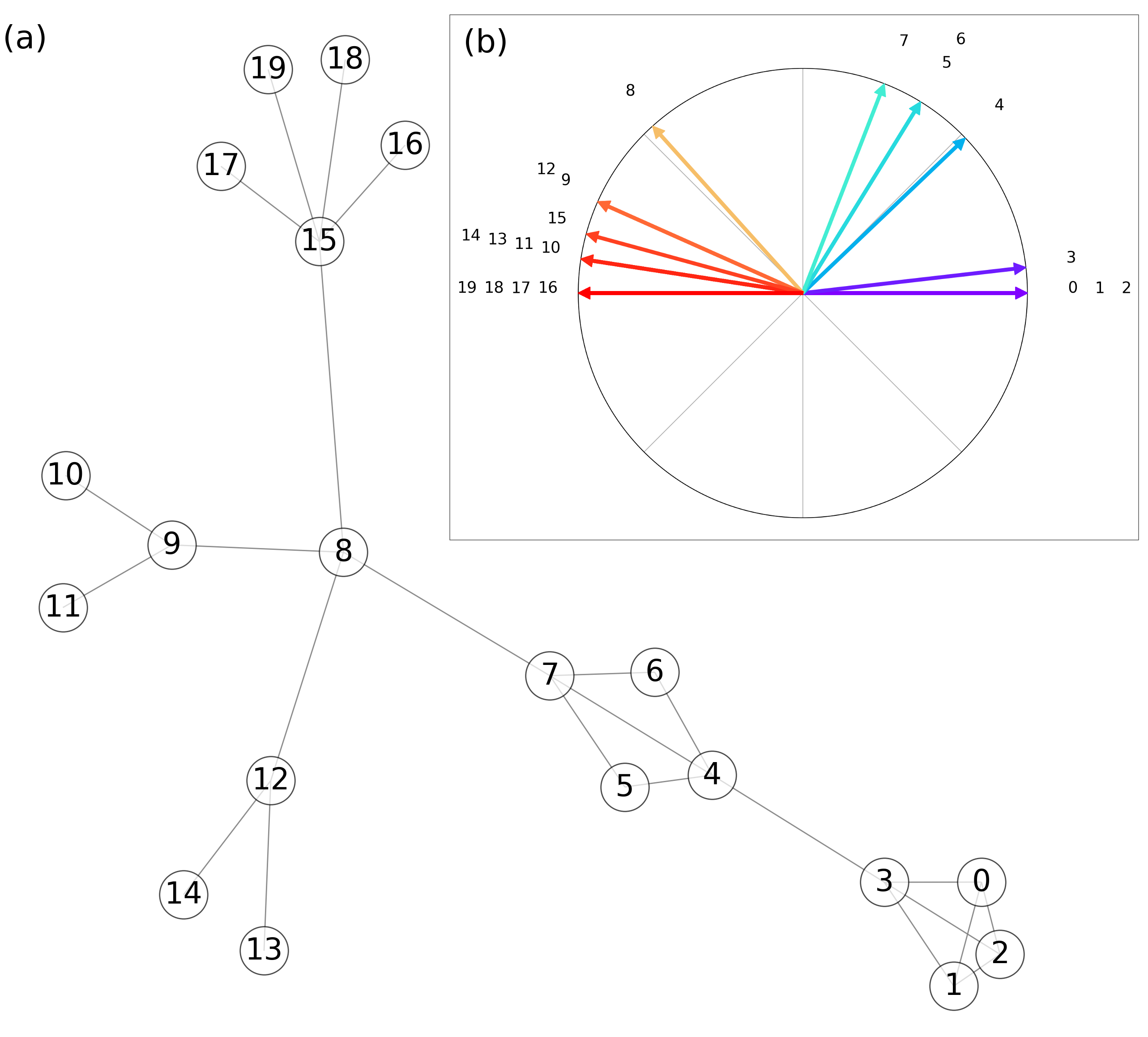}
\caption{(a) Labels' choice for the network of 20 nodes further examined in Figs \ref{fig:geometricDecomposition} and \ref{fig:structureQ} and its corresponding polar plot (b) of the $\{\phi_i\}$ scaled phases (in the range $[0,\pi]$) obtained according to Eq.(\ref{eq:phasesSolution}).}
\label{fig:structure}
\end{figure}

As $\alpha$ behaves as a scaling factor in Eq.(\ref{eq:phasesSolution}) one could always normalize the results such that $\phi_i \in [0, \pi]$. As an example, the values of $\vec{\phi}$, choosing $R=0$, for the network in Fig. \ref{fig:structure}(a) are 
\begin{eqnarray*}
\vec{\phi}=(0.0,0.0,0.0, 0.11,0.76,1.02,1.02,1.20, 2.30,2.72,2.99,2.99,2.72,2.99,2.99, 2.88,\\3.14,3.14,3.14,3.14)
\end{eqnarray*}
The corresponding clusters or orbits of the scaled values can be easily identified in the polar plot shown in Fig. \ref{fig:structure}(b).
Notice that the obtained groups are the same as the orbits coloured in Fig. \ref{fig:geometricDecomposition}(a), as expected.
\section{Quasi-symmetries in complex networks}
\label{sec:Quasi-symmetries}
The concept of approximate symmetry is not new. Approximate symmetry detection for 3D geometry \cite{Mitra2006} or approximate symmetry methods for solving differential equations \cite{Pakdemirli2004} are some examples. We address the question of what do we understand by approximate symmetries, or what we call \textit{quasi-symmetries}, in complex networks and how do they emerge. For that purpose, we will establish a simile with a circle, a geometric shape consisting of all points in a plane that are a constant distance, the radius, from the center. The circle is highly symmetric as every line that passes through the center generates a reflection and every angle represents a rotational symmetry around the center. However, one could obtain slightly different shapes if the points are obtained experimentally. Despite the underlying true shape being a circle, owing to missing data or experimental errors, the derived shape may lead to a deformed circle or \textit{quasi-symmetric} circle. Similarly, besides synthetic regular networks, real-world networks represent samples of processes that generate them and they are gathered by data collecting methods, either computational or experimentally. Ultimately, researchers deal with networks with missing or additional edges or nodes, as well as with noisy weighted networks. Hence, despite a group of nodes being structurally indistinguishable up to an error, that is, belonging to the same orbit, they may remain as separate independent units by applying traditional symmetry detection methods. 

As defined in Section \ref{sec:Symmetries}, the extent of symmetry of a symmetric graph can be measured by the number of possible symmetric permutations of its group of automorphisms \cite{Erdos1963}. We are concerned, however, by symmetry as a node-wise attribute. \textit{Symmetry,} as a mathematical concept, is a binary attribute of a node with respect to another, either true or false, depending on whether they belong to the same orbit or not. We however introduce the concept of \textit{quasi-symmetry} as a continuous variable that characterizes the degree of structural similarity of a pair of nodes. Obviously, a pair or a group of nodes that belong to the same orbit will be perfectly symmetric and therefore, have the largest possible value of quasi-symmetry. This new attribute enables us to characterize the degree of symmetry of all pair of nodes and provides richer information of the network. Notice that the concept of \textit{quasi-symmetry} can be applied not only to networks which have been perturbed, but also to networks of which we want to obtain the degree of symmetry between its nodes, even if we know, beforehand, that they do not belong to the same orbit.

Other authors have defined the notion of `near symmetry', a more restrictive definition of approximate symmetry, present in complex networks when certain properties remain invariant under some other network transformation, for example, node degree. Accordingly, notions of `stochastic symmetry' have also been established in order to characterize near symmetries in real networks \cite{Garlaschelli2010,Smith2019}.

\subsection{Building synthetic networks with quasi-symmetries}
\label{subsec:BuildingQuasi-symmetries}
\begin{figure}[H]
    \centering
    \includegraphics[width=0.5\textwidth]{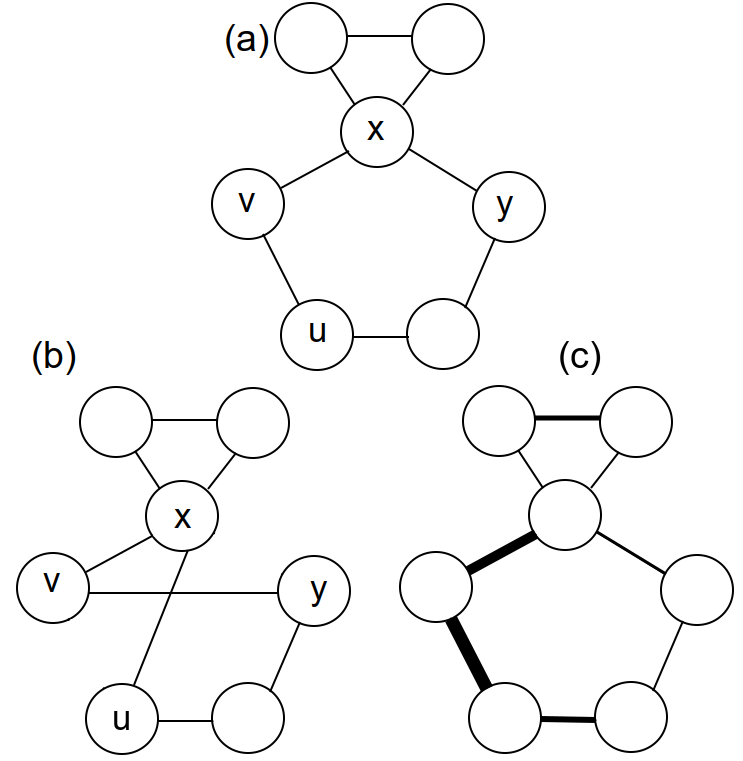}
    \caption{Example of a toy network with four orbits or clusters of equivalent nodes in panel (a), and two feasible quasi-symmetric networks drawing from it. Panel (b) shows one possible mechanism leading to the creation of quasi-symmetries by double-edge swapping and panel (c) exemplifies the perturbation of perfect symmetries by adding random weights to the edges.}
    \label{fig:createQuasiSymmetries}
\end{figure}

Real-world networks, both weighted and unweighted, are potential quasi-symmetric networks. In order to provide a general framework, we need to work with synthetic samples. As exposed in Section \ref{subsec:BuildingSymmetries}, random networks hardly present symmetric patterns and the latter are difficult to control. For this reason, we use the algorithm presented in Section \ref{subsec:BuildingSymmetries} in order to generate networks with any desired symmetry pattern. This networks are considered to be the underlying perfectly symmetric networks. On top of them, we build the quasi-symmetric networks by either swapping a given number of edges randomly or by modifying the weight of its edges. These mechanisms can be applied in very different ways. We present two particular implementations that can be applied in order to perturb the original networks. The first class of synthetic (unweighted) quasi-symmetric networks is constructed by swapping a random pair of edges, $\{(x,y),(u,v)\}$, that become $\{(x,u),(y,v)\}$ such that degree is preserved and the new edges do not already exist (See Fig. \ref{fig:createQuasiSymmetries}(b) for an example). The second class of (weighed) synthetic quasi-symmetric networks is constructed by adding a uniform random real number $w\in \mathcal{U}(-w_{max},w_{max})$ to the otherwise binary edge (See Fig. \ref{fig:createQuasiSymmetries}(c) for an example). The random transformations that result to a negative weight are ignored.
\subsection{Characterization of quasi-symmetries}
\label{subsec:characterization}
In Section \ref{subsec:SymmetriesDetection} we propose an alternative methodology to those based on discrete algebra for detecting the clusters of equivalent nodes or orbits of the network by bundling the nodes that have the same value of $\phi_i$ computed analytically from Eq.(\ref{eq:phasesSolution}). 
Using the same result, we extend the notion of symmetry into that of quasi-symmetry to characterize the degree of structural equivalence of all pair of nodes. 

We first compute the steady state phases of the nodes with respect to any reference node (we note that results do not depend on this choice) using Eq.(\ref{eq:phasesSolution}). The $\alpha$ parameter in Eq.(\ref{eq:phasesSolution}) acts as a scaling factor and hence one could always re-scale the set of phases such that they fall in the range $[0, \pi]$. In this way, the most distant nodes are separated by, at most, $\pi$ (See Fig. \ref{fig:structure}(b) as an example) and results are independent of the network size and the number of edges.

Next, the phase difference is computed between all pair of nodes as 
\begin{equation}
    \label{eq:phaseDifference}
    \Delta \phi_{ij} = |\phi_i-\phi_j| =  |\phi_j-\phi_i|
\end{equation}
Notice that $\Delta \phi_{ij} = 0$ if nodes $i$ and $j$ are completely symmetric. 

\subsubsection{Distribution of quasi-symmetries.}
\label{subsubsec:degree}
One could easily count the number of distinct orbits of a network with perfect symmetries either using a discrete algebra software or following the steps described in Section \ref{subsec:SymmetriesDetection}. But besides quantifying perfect symmetries, we may be interested in characterizing the topology of a network, regarding the structural similarity between the nodes, or quasi-symmetries. The first measure that we propose corresponds to the distribution of the scaled phases and phase differences.

In order to obtain more information about a network and distinguish whether it presents more structurally equivalent or similar nodes (quasi-symmetries) than those expected by a random network, we study two baseline types of networks and their distributions of quasi-symmetries: regular networks and random networks.
\begin{itemize}
    \item Regular Networks
    \begin{enumerate}
        \item Complete Networks $K_N$: all nodes are structurally equivalent, that is, they belong to the same orbit and, accordingly, they have the same value of $\phi_i$. Hence, $\Delta\phi_{ij}=0$ for all pairs of nodes. 
        \item Circulant Networks $G(k_1,...,k_m)$: in a circulant network, each node $i$ is connected to the nodes with indexes $i+k_s$ and $i-k_s$, for all the set of $m$ numbers. Many well-known graph families are subfamilies of the circulant networks. For example if $m=1$ and $k_1=1$, the resulting network is a circular network. The resulting distributions are delta-like, as for a complete network, as all nodes are structurally equivalent. 
        \item Balanced Tree Networks $G(r,h)$: A tree with a branching factor of $r$ and a height of $h$ has $\sum_{k=0}^{h}r^{k}=\frac{r^{h+1}-1}{r-1}$ nodes. The number of perfect symmetries or distinct orbits of the network is $h+1$, with a size given by $r^k$, where $k$ is the current height of the leaf. Therefore, there are $h(h+1)/2+1$ different values of $\Delta\phi_{ij}$, each one having $r^{k_1}r^{k_2}$ repetitions, where $k_1$ and $k_2$ are the height of the two leaves which we are considering. The frequency of $\Delta_{ij}=0$ corresponds to the count of all possible pairs of nodes in the same leaf, i.e., $\sum_{k=1}^{h}{r^k(r^k-1)}/2= \displaystyle \frac{r(r^h-1)(r^{h+1}-1)}{2(r^2-1)}$. Figure \ref{fig:treeDistribution} shows the distributions of scaled phases and the corresponding phase differences of a balanced tree network with a height of $h=3$ and two values of the branching factor,  $ r\in \{2,3 \}$. Notice that there are four distinct values of phases, according to $h+1=3+1$, with frequency given by $r^k$: $\{1,2,4,8\}$ and $\{1,3,9,27\}$, for $r=2$ and $r=3$, respectively (see Fig. \ref{fig:treeDistribution}(a,b). There are seven distinct values of phase differences, according to $h(h+1)/2+1 = 3(3+1)/2+1$ (see Fig. \ref{fig:treeDistribution}(c,d)).
    \end{enumerate}
\item Random Networks
    \begin{enumerate}
        \item Erd\"{o}s-R\'{e}nyi (ER) Network $G(N,p)$: in this model, each of the $\binom{N}{2}$ possible edges is included with probability $p$, independently from every other edge. Figure \ref{fig:ERdistribution} shows the distributions (relative frequencies) of the scaled phases, $\{\phi_{i}\}$ and the phase differences between nodes, $\{\Delta\phi_{ij}\}$ for an ER network of 500 nodes and three different values of the $p$. As the probability of connection approaches $1$, the network becomes closer to a complete network and therefore, there are more nodes that are structurally similar. Consequently, the distribution of scaled phases and phase differences is discrete (see the bottom panels in Fig.  \ref{fig:ERdistribution}(a-b)). Intermediate values of $p$ lead to a continuous distribution of scaled phases which average approaches $\pi/2$ as $p$ increases (see the middle panels in Fig. \ref{fig:ERdistribution}(a-b)). The final shape of the distribution is a reflection of the degree distribution of the original network.
     \item Barab\'{a}si-Albert (BA) Network $G(N,m)$: in this model, called preferential attachment or Barab\'{a}si-Albert network, nodes are added one at a time with $m$ random edges which are linked to the existing nodes with a probability proportional to the degree of them. 
Figure \ref{fig:BAdistribution} shows the distributions (relative frequencies) of the scaled phases, $\{\phi_{i}\}$ and the phase differences between nodes, $\{\Delta\phi_{ij}\}$ for a BA network of 500 nodes and three different values of the $m$. The resulting distributions are very similar to that of ER networks (see Fig. \ref{fig:ERdistribution}). Besides small values of $m$, resulting to star-like patterns, the distribution of phases is continuous. Again, the particular shape of the distributions is determined by the degree distribution of the original network. 
    \end{enumerate}

\begin{figure}[H]
		\centering
		\includegraphics[width=\textwidth]{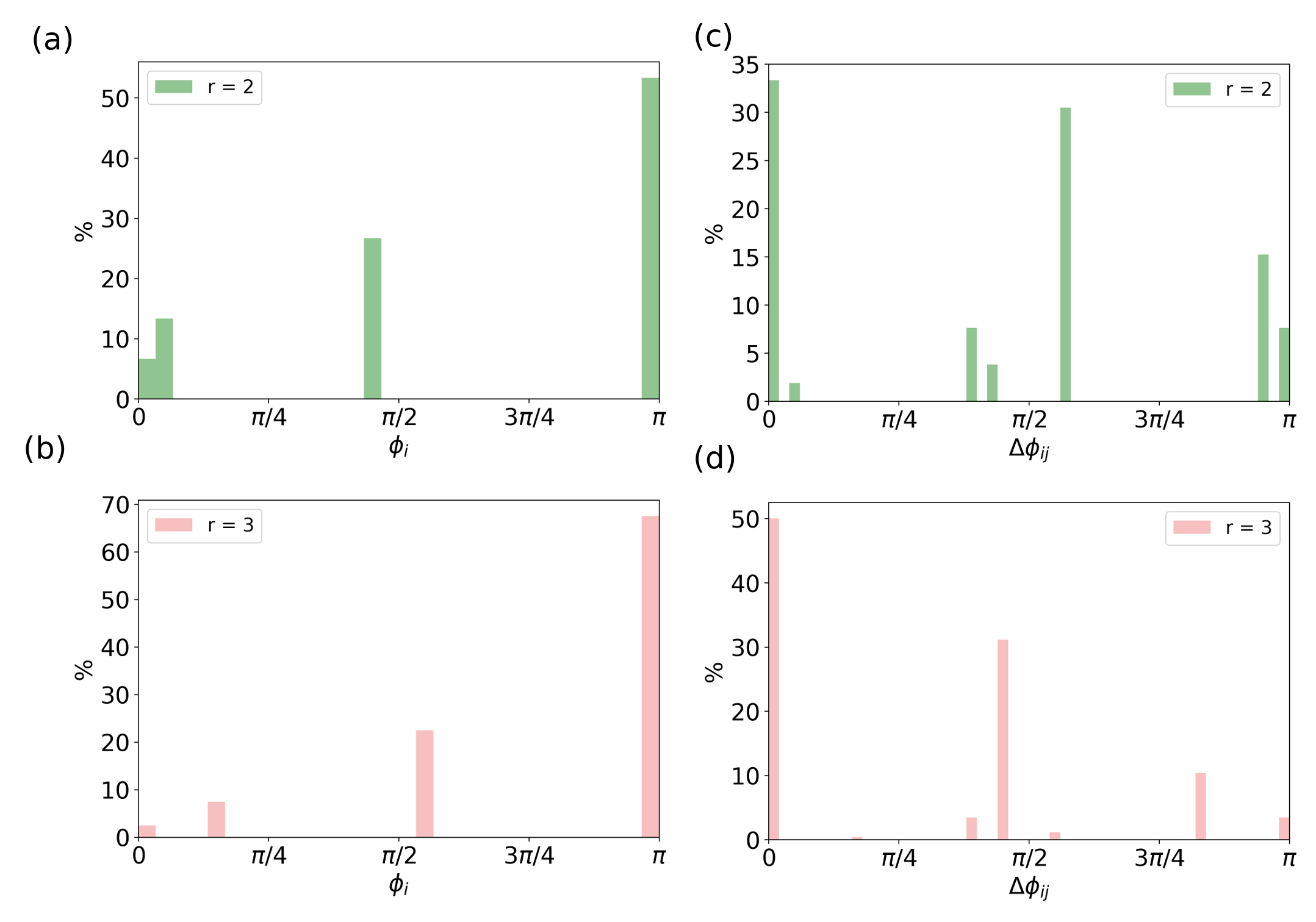}
		\caption{Relative frequency of the scaled phases, $\phi_i$, obtained using Eq.(\ref{eq:phasesSolution}), and phase differences, $\Delta \phi_{ij}$, of a balanced tree network of height, $h$, equal to 3 and branching factor, $r$, of $2$ [panels (a) and (c)] and $3$ [panels (b) and (d)].}
		\label{fig:treeDistribution}
	\end{figure}  
\begin{figure}[H]
		\centering
		\includegraphics[width=\textwidth]{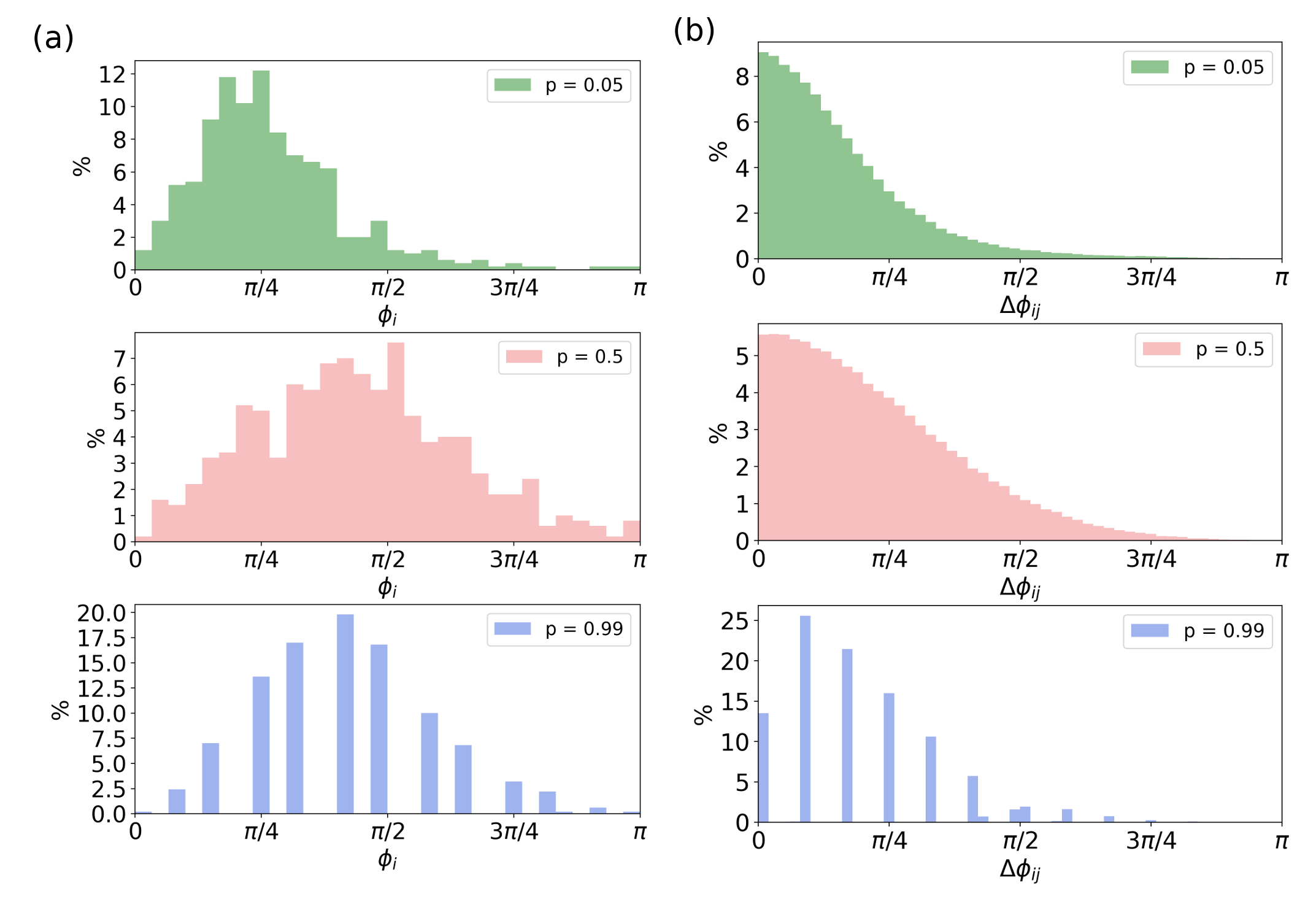}
		\caption{Relative frequency of the scaled phases, $\phi_i$, obtained using Eq.(\ref{eq:phasesSolution}), and phase differences, $\Delta \phi_{ij}$, of an Erd\"{o}s-R\'{e}nyi random network of 500 nodes and three different densities: $p=0.05$, $p=0.5$ and $p=0.99$ (upper, middle and lower figures in panels (a) and (b), respectively).}
		\label{fig:ERdistribution}
	\end{figure} 
\begin{figure}[H]
		\centering
		\includegraphics[width=\textwidth]{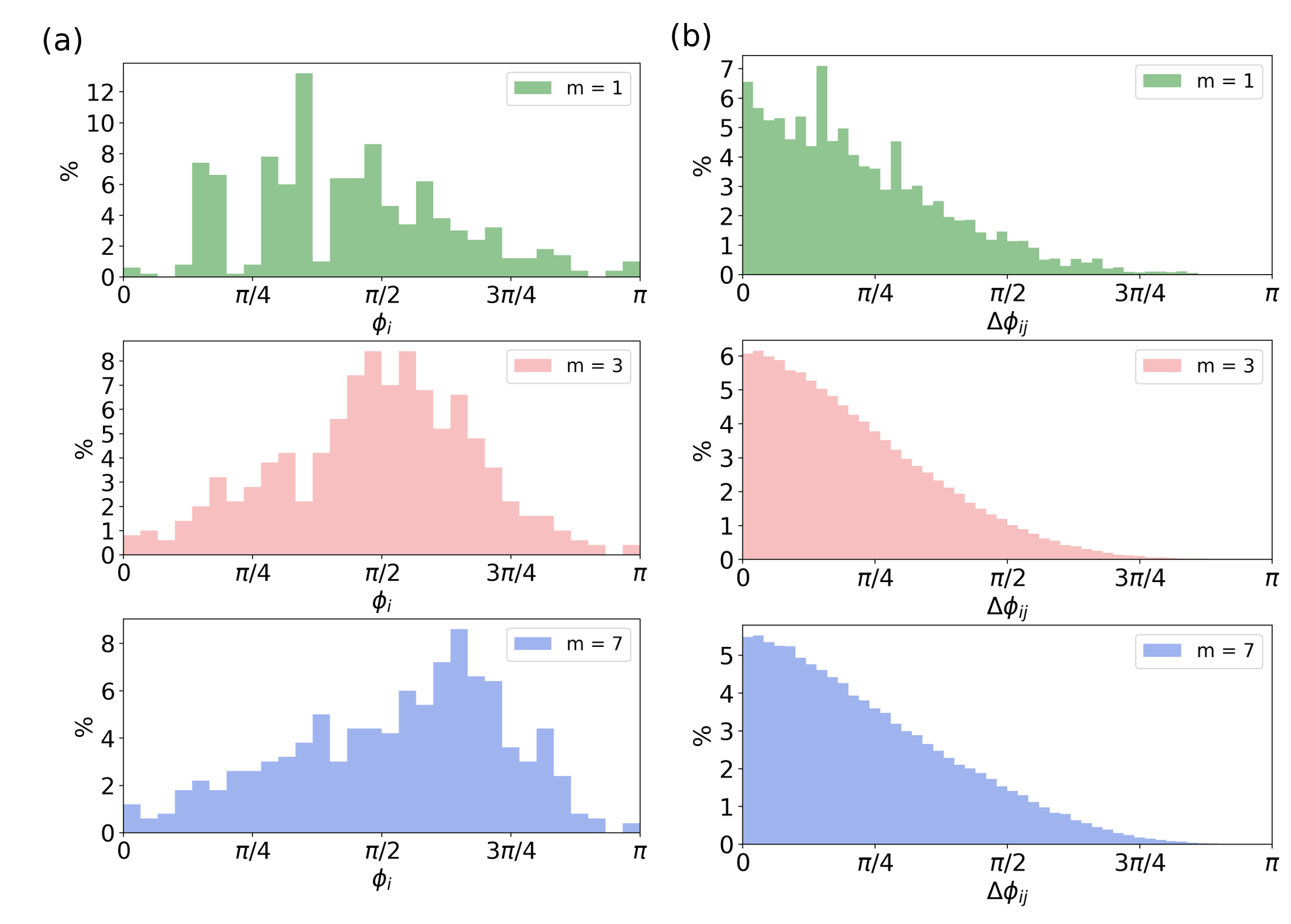}
		\caption{Relative frequency of the scaled phases, $\phi_i$, obtained using Eq.(\ref{eq:phasesSolution}), and phase differences, $\Delta \phi_{ij}$, of a Barab\'{a}si-Albert random network of 500 nodes and three different densities: $m=1$, $m=3$ and $m=7$ (upper, middle and lower figures in panels (a) and (b), respectively).}
		\label{fig:BAdistribution}
\end{figure}  
From the analysis of random networks, namely, ER and BA models, we conclude that, despite both networks have distinct network topologies, i.e., different degree distributions, the level of structural similarity between nodes is very similar. We conclude that random networks display a uni-modal continuous distribution of phases, the shape of which is determined by the corresponding degree distribution. Extreme values of the parameters of the models, i.e., very few connections or a large value of the density, conversely, lead to a discrete distribution of phases, resulting from most of the nodes being structurally similar.
\item Networks with perturbed (quasi) symmetries

Once regular and random networks have been analysed in terms of structure similarity, we conduct the equivalent analysis of networks of which we control the number of (perfect) symmetries, generated accordingly to the methodology presented in Section \ref{subsec:BuildingSymmetries}. In order to assess the effect of perturbing the originally perfect symmetries, we add a uniform random noise to each edge (see Section \ref{subsec:BuildingQuasi-symmetries}) and study the evolution of similarity, or presence of \textit{quasi-symmetries}, of two networks with $5$ and $12$ symmetries in the non-perturbed network (originally perfect symmetries). Figs \ref{fig:5symmetriesDistribution}(a) and \ref{fig:12symmetriesDistribution}(a) show the relative frequency of the scaled phases, $\phi_i$, obtained using Eq.(\ref{eq:phasesSolution}) of a network of 264 nodes with originally 5 perfect symmetries or orbits and another of 209 nodes and originally 12 perfect symmetries, respectively. Six different perturbed networks are included for each one, by adding a uniform random noise in the range $[-w_{max},w_{max}]$ to each edge.
\begin{figure}[H]
		\centering
		\includegraphics[width=\textwidth]{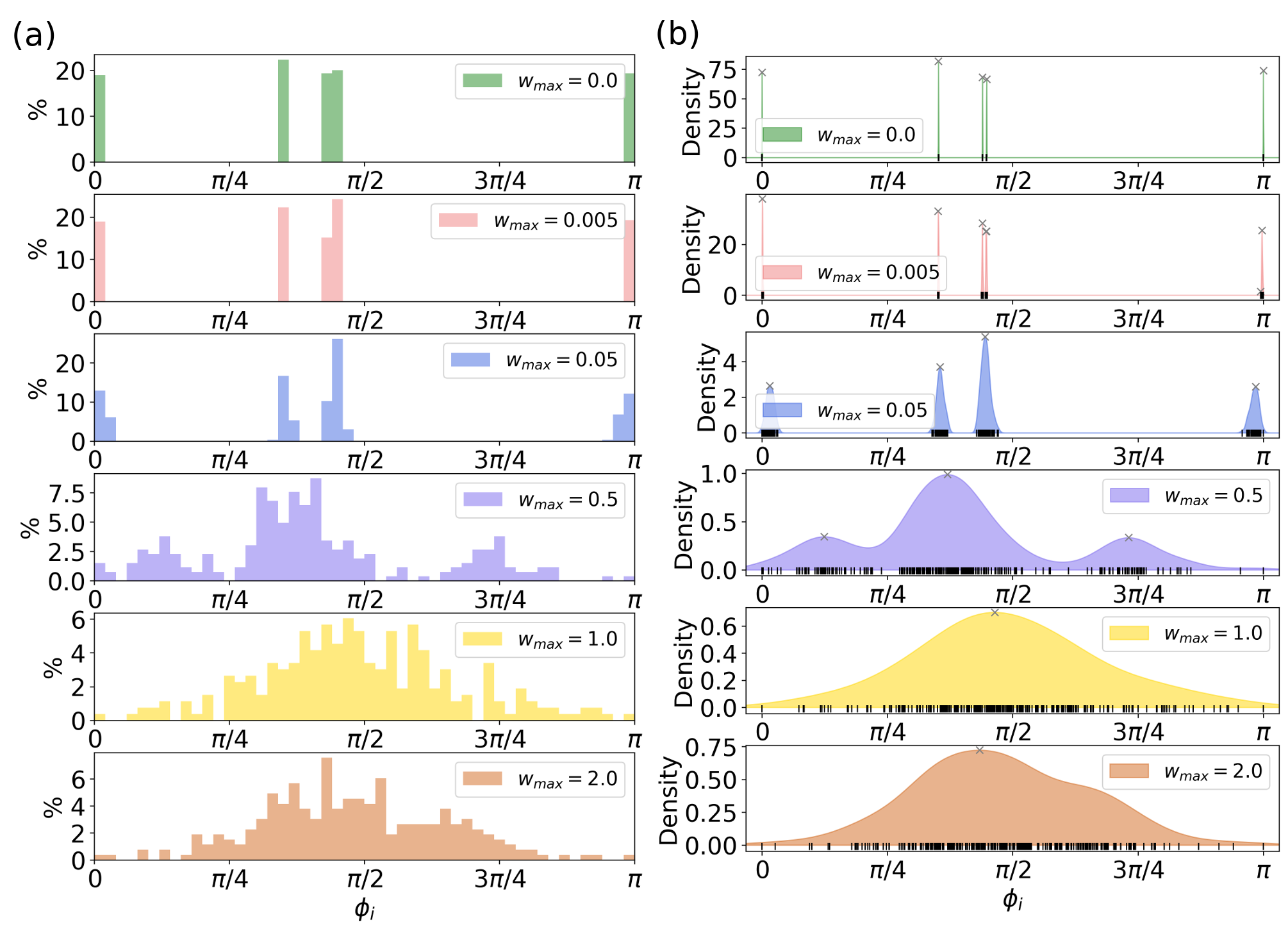}
		\caption{Relative frequency of the scaled phases, $\phi_i$ [panel (a)], obtained using Eq.(\ref{eq:phasesSolution}) and the corresponding Gaussian Kernel Density distributions with an optimal bandwidth choice according to cross-validation method of a network of 264 nodes with originally 5 perfect symmetries or orbits. A uniform random noise in the range $[-w_{max},w_{max}]$ is added to each edge, avoiding negative values. Six different values of $w_{max}$ are included, from upper to lower panels.}
		\label{fig:5symmetriesDistribution}
\end{figure}
\begin{figure}[H]
		\centering
		\includegraphics[width=\textwidth]{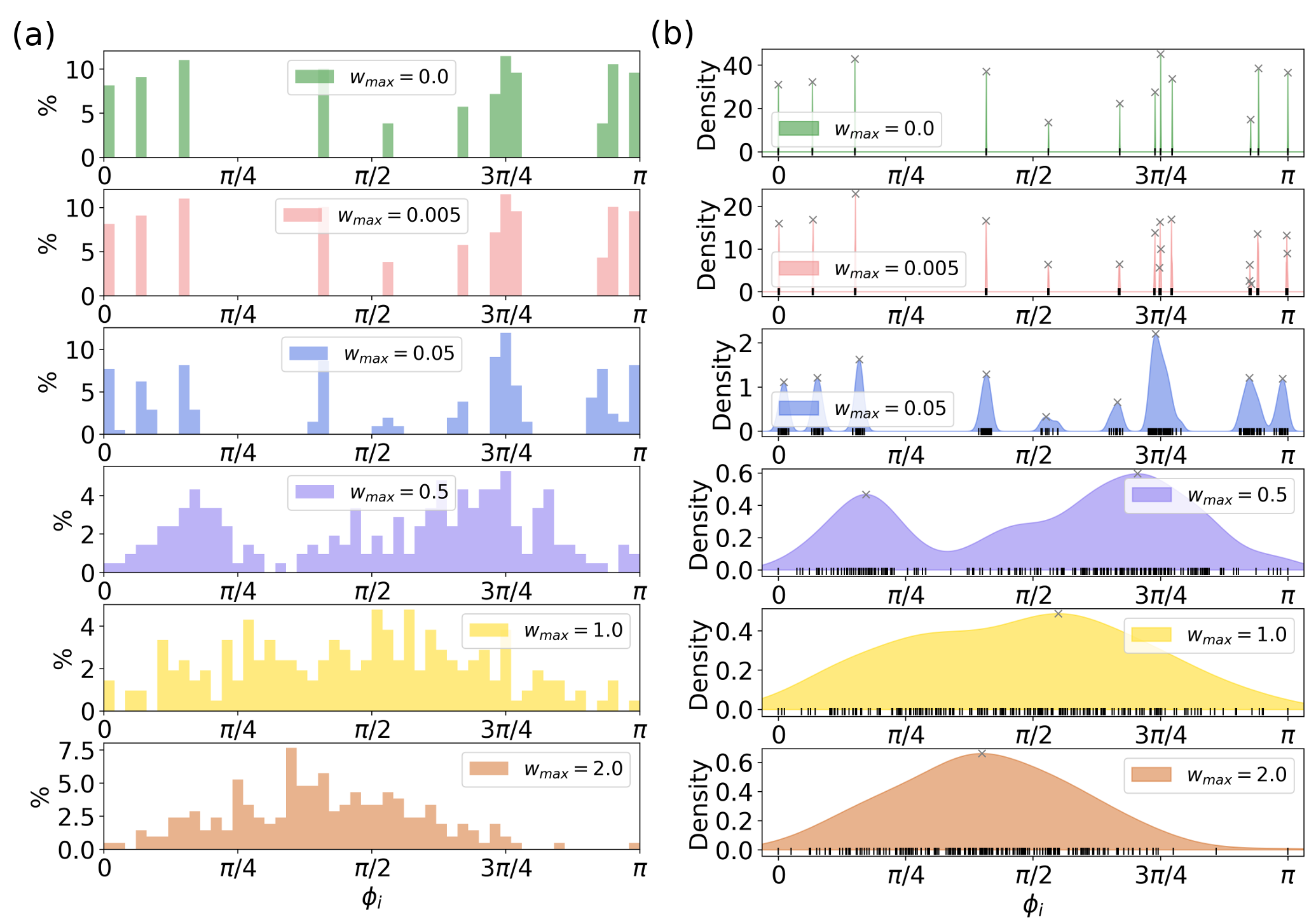}
		\caption{Relative frequency of the scaled phases, $\phi_i$ [panel (a)], obtained using Eq.(\ref{eq:phasesSolution}) and the corresponding Gaussian Kernel Density distributions with an optimal bandwidth choice according to cross-validation method of a network of 209 nodes with originally 12 perfect symmetries or orbits. A uniform random noise in the range $[-w_{max},w_{max}]$ is added to each edge, avoiding negative values. Six different values of $w_{max}$ are included, from upper to lower panels.}
		\label{fig:12symmetriesDistribution}
\end{figure}
The upper panels in Figs \ref{fig:5symmetriesDistribution}(a) and \ref{fig:12symmetriesDistribution}(a) show that, in the non-perturbed case, i.e., $w_{max}=0.0$, the distribution of scaled phases results to a discrete one that sets the group of nodes apart according to symmetries, $5$ and $12$, respectively. As the network becomes more noisy, i.e, the value of $w_{max}$ increases, the symmetries are no longer perfect, but \textit{quasi-symmetries}. In other words, equivalent nodes turn to similar nodes. When the perturbation applied to the networks is too large, the distributions of phases is similar to that of a random network (see lower panels in Figs \ref{fig:5symmetriesDistribution}(a) and \ref{fig:12symmetriesDistribution}(a)).

Thereby, we have shown that networks which structure is enriched with quasi-symmetries, differently than random networks, present a very particular pattern regarding phases distribution, even if perfect symmetries have been removed by adding random noise: they are characterized by a multi-modal distribution of phases, rather than the uni-modal distribution that identify random networks.

\subsubsection{Counting quasi-symmetries.}
\label{subsubsec:countingQuasiSymmetries}
In order to assess the quality of the \textit{quasi-symmetries} or structural similarity of real-world networks, we propose a methodology to reject the hypothesis that the network presents a structural similarity equivalent to that of a random network. To do so, we explore the modality of the scaled phases distribution. In other words, we detect the number of modes or peaks of the distribution of scaled phases. When the distribution of $\{\phi_i\}$ is uni-modal we can not say that the topology of the network is different to that of a random network, with respect to \textit{quasi-symmetries} or structural similarity.

The methodology consists in fitting a Gaussian Kernel Density Estimator (KDE) distribution to the scaled phases (see the Appendix \ref{KDE} for more details on KDE). The bandwidth of the Kernel, for each case, is selected using cross-validation, a non-parametric methodology. \cite{Hall1991, Taylor1989}. The results for the networks of $5$ and $12$ symmetries are shown in Figs \ref{fig:5symmetriesDistribution}(b) and \ref{fig:12symmetriesDistribution}(b), respectively. Notice that the width of the distributions changes with increasing $w_{max}$, as expected. Once the optimal density is found for each $w_{max}$, we can detect the peaks for each case. Notice that the perfect symmetries are unequivocally detected (see upper panels in Figs \ref{fig:5symmetriesDistribution}(b) and \ref{fig:12symmetriesDistribution}(b)). The distribution becomes more broadened and the number of detected peaks diminishes. Finally, when the networks are completely perturbed, i.e., randomized, the distributions and corresponding number of peaks, or modes, are equivalent to the random networks presented in Figs \ref{fig:ERdistribution} and \ref{fig:BAdistribution}, that is, uni-modal distributions.
\end{itemize}
From the analysis of the symmetric networks we can draw several conclusions, which will be applied to real-world networks: firstly and importantly, random networks present a uni-modal distribution of scaled phases. Secondly, networks that are made of groups of structurally similar nodes, present multi-modal distribution of scaled phases. Narrower peaks are a signal of more differentiated groups of nodes. 

\begin{figure}[H]
    \centering
    \includegraphics[width=0.55\textwidth]{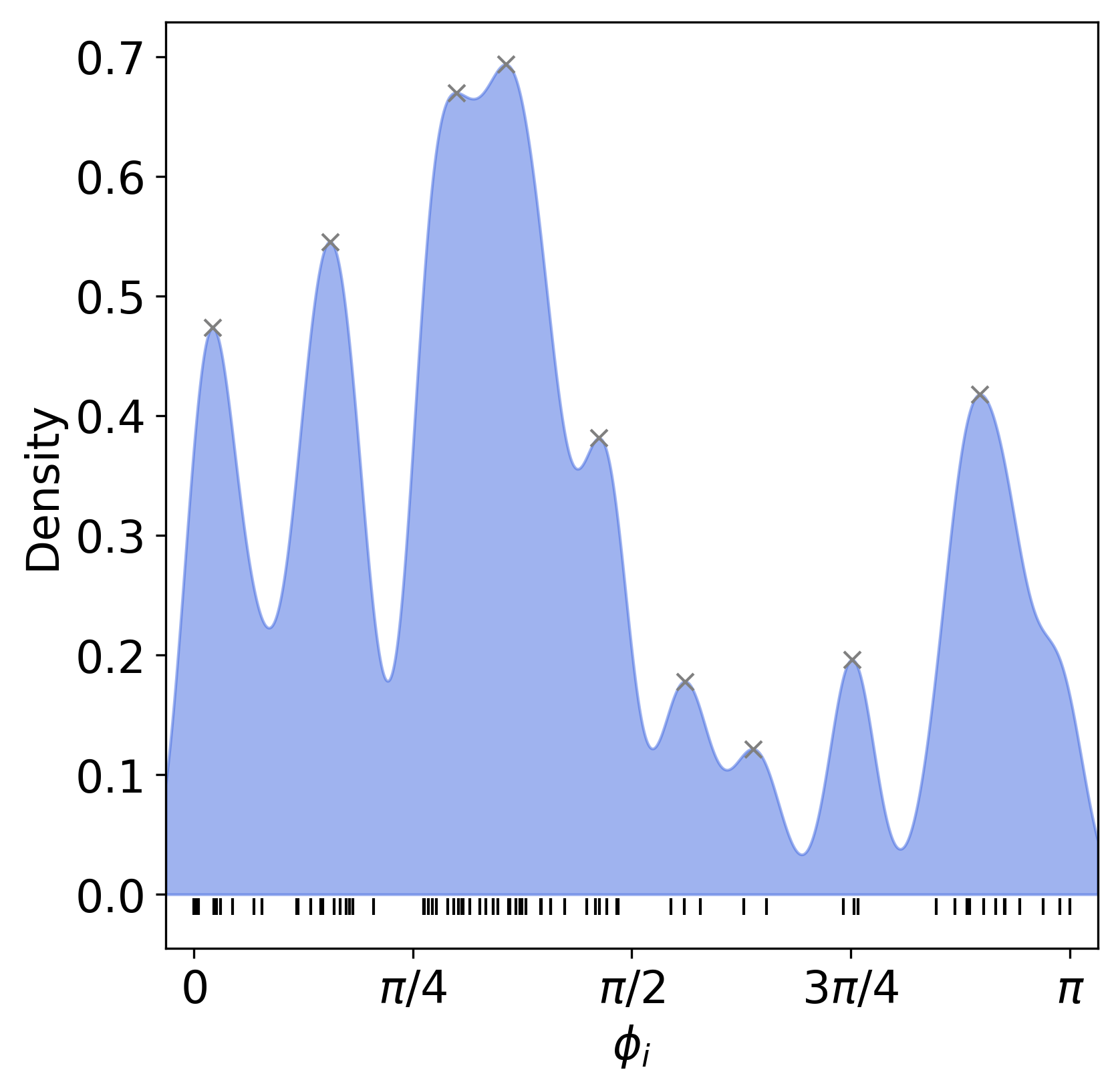}
    \caption{Kernel Density Distribution, using a Gaussian kernel, with an optimal choice of the bandwidth according to 10\% cross-validation method of a Macaque brain network, where the nodes above the 95\% percentile of phases are removed (outliers).}
    \label{fig:macaque_kde}
\end{figure}

In Fig. \ref{fig:macaque_kde} we show the KDE distribution, with the optimal choice of bandwidth, and the corresponding peaks, for a whole-cortex macaque structural connectome constructed from a combination of axonal tract-tracing and diffusion-weighted imaging data \cite{macaque2018}, which displays distinct modes and hence, informs us that the network is more richer than a random topology with regard to \textit{quasi-symmetries}.
\section{The Dual Network}
\label{sec:dualNetwork}
The analysis of the distribution of scaled phases and the corresponding phase differences leaves much room for obtaining more in-depth insights of the structural similarity or \textit{quasi-symmetries} in complex networks. Eq.(\ref{eq:phaseDifference}) enables us to define the \textit{dual network}, a mathematical object which gathers all the information regarding the \textit{quasi-symmetries} of a network, as we will see.

We define the dual network, $\mathcal{H}(\mathcal{V}, \mathcal{E'})$ of  $\mathcal{G(\mathcal{V}, \mathcal{E})}$, as a complete weighted network that inherits all nodes of the original one and which edges are given a weight according to
\begin{equation}
    \label{eq:weightsDual}
    w_{ij} = \frac{\pi-\Delta\phi_{ij}}{\pi}
\end{equation}Hence, the weight of the edges is in the range $[0,1]$. An edge connecting two nodes which are completely symmetric has a weight of $1$, while an edge connecting the most distant nodes has a weight of $0$. Notice that weights are obtained from phase differences applying a linear transformation.
\begin{figure}[H]
\centering
\includegraphics[width=\textwidth]{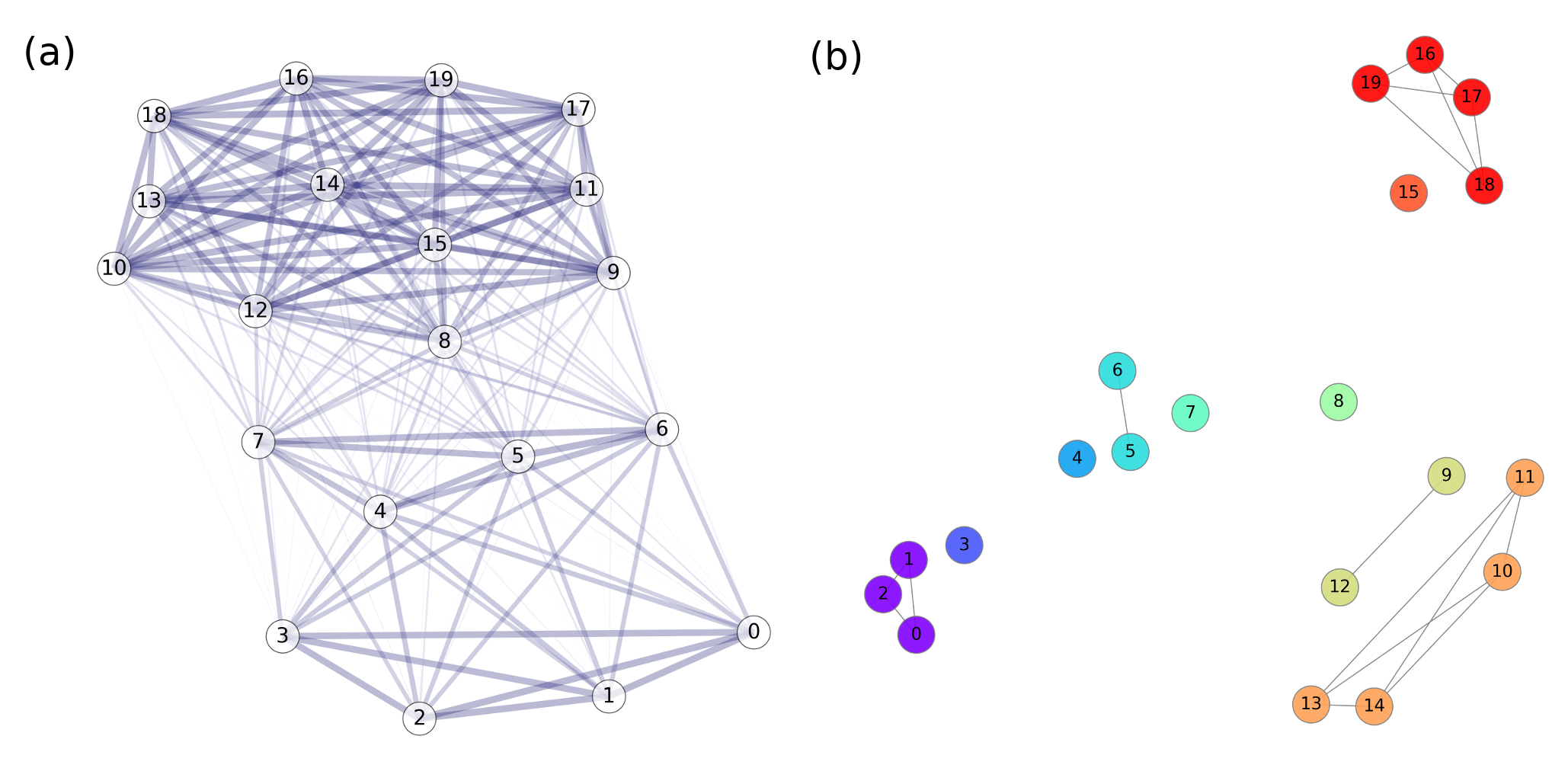}
\caption{(a) Dual network, $\mathcal{H}$, corresponding to the network, $\mathcal{G}$, in Fig. \ref{fig:structure}(a) with its phases distributed as shown in Fig. \ref{fig:structure}(a). Edge width and color intensity scales with its weight, computed using Eq.(\ref{eq:weightsDual}). The position of nodes are computed using the Fruchterman-Reingold force-directed algorithm \cite{Fruchterman1991} considering $\mathcal{H}$. (b) Corresponding binarized dual network,  $\mathcal{H}_{\mathcal{B}}$. The number of communities (in different colors) is set to $10$, the known value of different orbits. The position of nodes are computed using the Fruchterman-Reingold force-directed algorithm \cite{Fruchterman1991} considering $\mathcal{G}$.}
\label{fig:structureDual}
\end{figure}

Figure \ref{fig:structureDual}(a) shows the dual network corresponding to the network in Fig. \ref{fig:structure}(a) with its phases distributed as shown in Fig. \ref{fig:structure}(b). Notice that the nodes which are structurally more similar are more strongly connected, i.e., the edges connecting them have a larger weight, and they are placed very close together when using the Fruchterman-Reingold force-directed algorithm \cite{Fruchterman1991} for assigning the position of the nodes in the layout of the network. In the network of Fig. \ref{fig:structure}(a), many nodes are part of tree-like motifs of different sizes. This structural similarity is reflected in them being tightly connected in the corresponding dual network.

The fact that the dual network, it is worth saying, is a complex network, entails that many measures developed in the field of network theory can be also applied to this particular network, unveiling interesting properties of the original one. 

Before exploring the most informative measures on the dual network, we define the binnarized dual network, $\mathcal{H}_{\mathcal{B}}$, as the network with the adjacency matrix given by
\begin{equation}
    \label{eq:binnaryDual}
    a_{ij} = \begin{cases}
    1 \text{\hspace{1.5cm}if } w_{ij} \geq w_{threshold} \\
    0 \text{\hspace{1.5cm}otherwise}  \\
    \end{cases}
\end{equation}
In other words, a threshold value for the weight determines the sparsity of the binnarized dual network. $\mathcal{H}_{\mathcal{B}}$ leads to more significant results of network measures, as explained in Section \ref{subsec:centrality}. Different values of $w_{threshold}$ enhance or weaken the presence of \textit{quasi-symmetries}, ranging from a complete network to a completely disconnected one. Our approach consists in choosing a threshold such that the corresponding number of detected communities in $\mathcal{H}_{\mathcal{B}}$ is the same as the number of peaks obtained in the Gaussian Kernel density. Note that several values may verify the latter requirement, a fact that captures the probabilistic nature of a network with \textit{quasi-symmetries}. As long as the main edges are conserved, one could obtain the same number of communities with slightly different connectivity patterns. 
Figure \ref{fig:structureDual}(b) shows the binnarized dual network, $\mathcal{H}_{\mathcal{B}}$, corresponding to the network in Fig. \ref{fig:structure}(a). The number of communities, in this case, perfect symmetries, is $10$, and $w_{threshold}$ is chosen to meet this requirement. Notice that only nodes that belong to the same orbit are connected by an edge, while all nodes remain connected in the original definition of (weighted) dual network (see Fig. \ref{fig:structureDual}(a)). Note that different values of $w_{threshold}$ may lead to the same number of communities. This behaviour becomes more clear when dealing with larger networks. If we consider the networks with originally $5$ and $12$ symmetries when we apply a perturbation on their edges with $w_{max}=0.05$, the number of detected peaks is $4$ and $9$, respectively (see Figs \ref{fig:5symmetriesDistribution}(b) and \ref{fig:12symmetriesDistribution}(b)). Using Eq.(\ref{eq:binnaryDual}), we obtain the corresponding $\mathcal{H}_{\mathcal{B}}$, by setting the number of communities to the number of detected peaks. A range of $w_{threshold}$ values leads to feasible networks and one can choose between more sparse networks (higher values of $w_{threshold}$) or more densely connected (lower values of $w_{threshold}$) realizations of $\mathcal{H}_{\mathcal{B}}$. 
\begin{figure}[H]
\centering
\includegraphics[width=\textwidth]{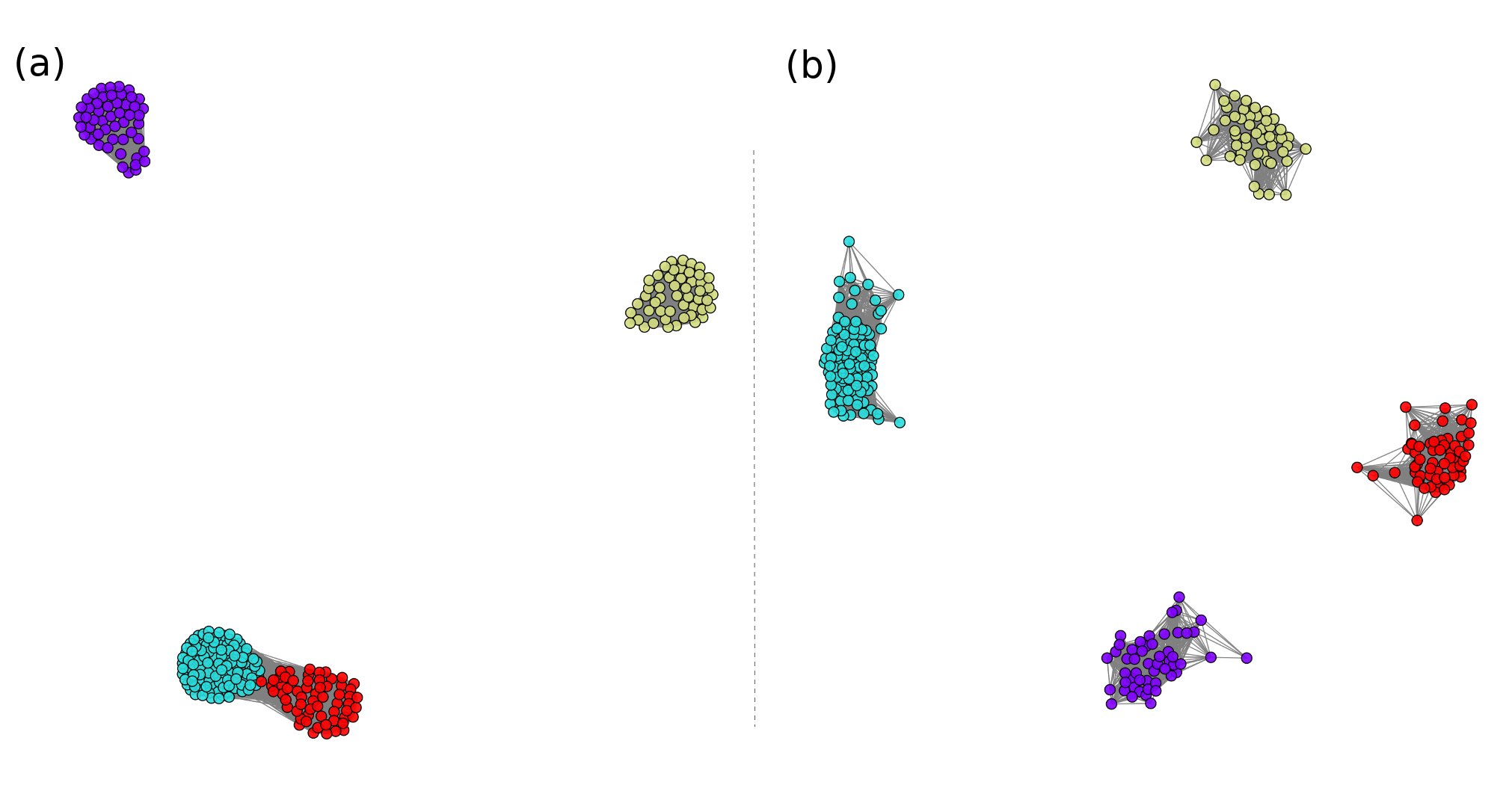}
\caption{$\mathcal{H}_{\mathcal{B}}$ for the network with originally $5$ symmetries when we apply a perturbation with $w_{max}=0.05$, corresponding to $4$ peaks in the distribution of scaled phases (see Section \ref{subsec:BuildingQuasi-symmetries}). The position of nodes are computed using the Fruchterman-Reingold force-directed algorithm \cite{Fruchterman1991} considering $\mathcal{H}_{\mathcal{B}}$. The threshold values for the weight are $w_{threshold}=0.92$ and $w_{threshold}=0.99$, in panel (a) and (b), respectively.}
\label{fig:5symmetries_HB}
\end{figure}
\begin{figure}[H]
\centering
\includegraphics[width=\textwidth]{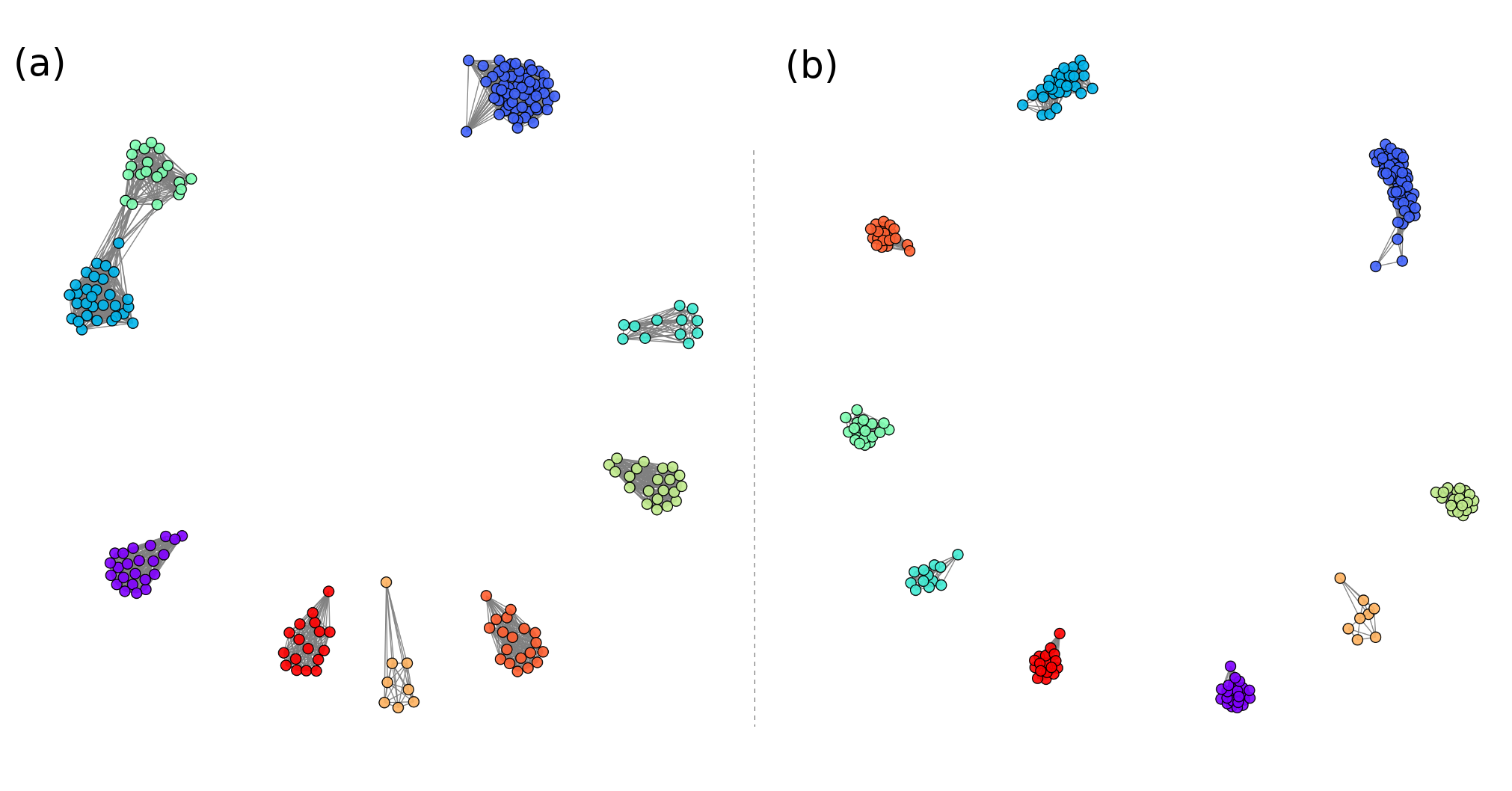}
\caption{$\mathcal{H}_{\mathcal{B}}$ for the network with originally $12$ symmetries when we apply a perturbation with $w_{max}=0.05$, corresponding to $9$ peaks in the distribution of scaled phases (see Section \ref{subsec:BuildingQuasi-symmetries}). The position of nodes are computed using the Fruchterman-Reingold force-directed algorithm \cite{Fruchterman1991} considering $\mathcal{H}_{\mathcal{B}}$. The threshold values for the weight are $w_{threshold}=0.96$ and $w_{threshold}=0.98$, in panel (a) and (b), respectively.}
\label{fig:12symmetries_HB}
\end{figure}
Figure \ref{fig:5symmetries_HB} shows two feasible $\mathcal{H}_{\mathcal{B}}$ for the network with originally $5$ symmetries after applying a perturbation with $w_{max}=0.05$ (see Section \ref{subsec:BuildingQuasi-symmetries}), corresponding to $4$ detected peaks in the distribution of scaled phases. Notice that when $w_{threshold}$ is larger, the dual network becomes more sparse (Fig. \ref{fig:5symmetries_HB}(b)). However, the number of communities is preserved, as a requirement for the creation of $\mathcal{H}_{\mathcal{B}}$. The same applies in Fig. \ref{fig:12symmetries_HB}, for the network with originally $12$ symmetries when we apply a perturbation with $w_{max}=0.05$, corresponding to $9$ detected peaks.

\subsection{Centrality measures.}
\label{subsec:centrality}
The characterization of the nodes in a network includes the study of its centrality, a measure of its importance in the network, based on the application-specific context we are interested in. On this basis, centrality measures are classified into two types, depending on whether local information around the particular nodes or global information of the network is required. 

At the beginning of this Section we have introduced the definition of the dual network, $\mathcal{H}$, and the corresponding binnarized network, $\mathcal{H}_{\mathcal{B}}$, which represents a mapping of the structural similarity between nodes of the original networks, namely, its quasi-symmetries. In fact, the dual network is the more complete measure of the structural similarity or equivalence between nodes, relying on the distribution of quasi-symmetries, as exposed in section \ref{subsubsec:degree}. Nonetheless, we may be interested in computing node-specific measures that inform us about the role that a particular node plays regarding the structural similarity of a network. To this end, we explore some well-known centrality measures on top of the dual network to obtain new insights about the nodes that are the most relevant concerning structural similarity. Although many centrality measures have been proposed, we focus on providing an analysis of one local and one global centrality measures, namely, degree and betweenness centralities, respectively. 

\begin{figure}[H]
    \centering
    \includegraphics[width=0.6\textwidth]{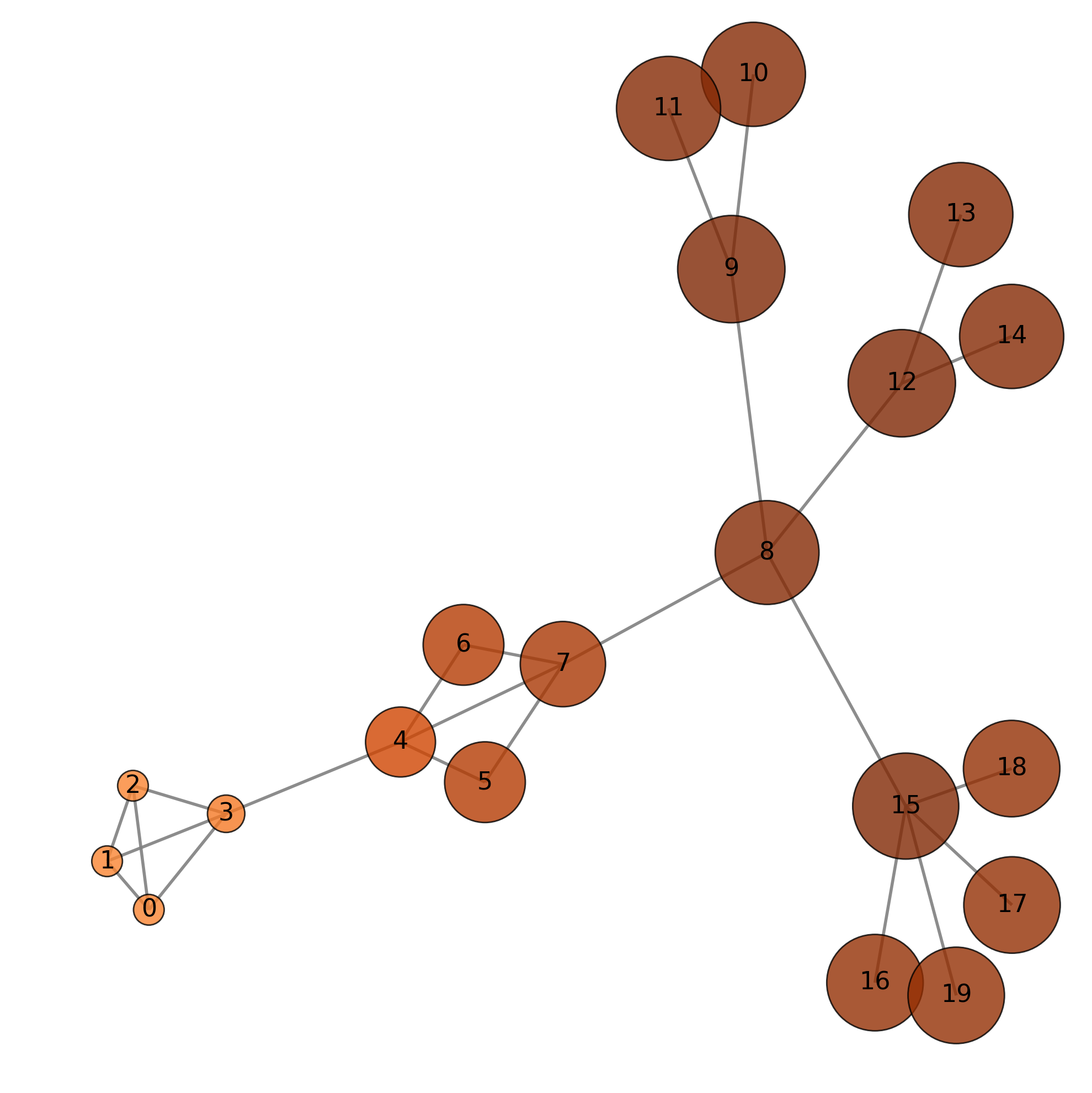}
    \caption{Network defined in Fig. \ref{fig:structure}(a), such that the radius of the nodes is proportional to the values of the degree centrality of the dual network, and the color map is built such that darker values correspond to larger values of this centrality. The position of nodes are computed using the Fruchterman-Reingold force-directed algorithm \cite{Fruchterman1991} considering the original network.}
    \label{fig:structure_DegreeOriginalLayout}
\end{figure}

We provide an example of the degree centrality values in $\mathcal{H}$ for the network defined in Fig. \ref{fig:structure}(a). The radius of the nodes in Fig. \ref{fig:structure_DegreeOriginalLayout} is proportional to the values of the degree centrality of the dual network, and the color map is built such that darker values correspond to larger values of this centrality. Nodes that have the largest values are those that are more structurally similar to all others, while nodes with the smallest values are those whose position is more rare or unique. 
\begin{table}[H]
\centering
\begin{tabular}{|c|c|} \hline
 Node ID & Degree Centrality \\  \hline
 0,1,2 & 0.342 \\   
 3 & 0.369 \\
 4 & 0.499 \\
 5,6 & 0.542 \\
 7 & 0.560 \\
 16, 17, 18, 19 & 0.605 \\
 8 & 0.634 \\
 10, 11, 13, 14 & 0.636 \\
 15 & 0.643 \\
 9, 12 & 0.648 \\ \hline
\end{tabular}
\label{tab:degreeCentrality} 
 \caption{Values of the degree centrality of the dual network corresponding to the network in Fig. \ref{fig:structure}(a) sorted in ascending order.}   
\end{table}
Table \ref{tab:degreeCentrality} shows the results of the degree centrality of the dual network corresponding to the network in Fig. \ref{fig:structure}(a) sorted in ascending order. The nodes that display a largest value of degree centrality in the dual network are the nodes $9$ and $12$, while those displaying the smallest values correspond to nodes $0, 1$ and $2$. 

In Fig. \ref{fig:macaque_Degree_NonSparse} we show the results of degree centrality on  $\mathcal{H}_{\mathcal{B}}$ for the Macaque brain network, which KDE distribution is presented in Fig. \ref{fig:macaque_kde}. Although we are not looking for the interpretation of the obtained results, as it is not our field of expertise, we highlight the fact that brain regions that display a larger value of degree centrality account for a larger similarity with many other nodes in the network (here we find dorsolatral premotor cortex, prefrontal polar cortex, superior parietal cortex, posterior insula and orbitolateral prefrontal cortex as the most central ones), while small values of degree centrality means that their role in the network is more unique, in the sense that no other nodes can play a similar structural role (here we find amygdala, inferior temporal cortex, primary visual cortex, anterior visual area, ventral temporal cortex and hippocampus as the less central ones).  

\begin{figure}[H]
    \centering
    \includegraphics[width=\textwidth]{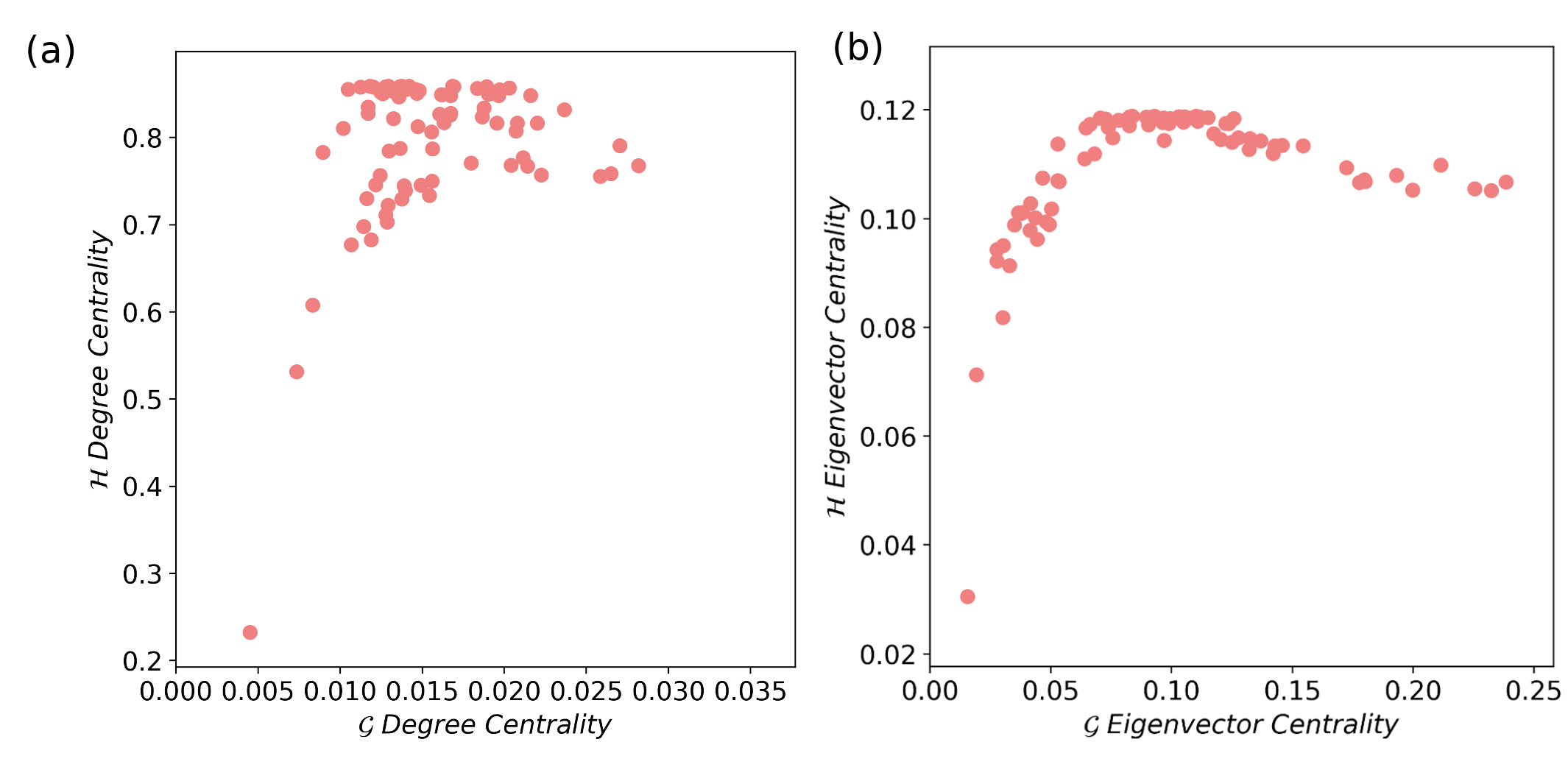}
    \caption{Scatter plot of the degree and eigenvector centralities, in panel (a) and (b), respectively, obtained for both the original and its corresponding dual network, $\mathcal{G}$ and $\mathcal{H}$, respectively, for the brain network of a Macaque.}
    \label{fig:macaqueCentralitiesCompare}
\end{figure}

Figure \ref{fig:macaqueCentralitiesCompare} shows for the Macaque brain network, the relation between the original network and its corresponding dual network regarding degree and eigenvector centralities, respectively. Note that the most central nodes of the original network and its dual counterpart are not the same. Therefore, the dual network provides new insights about the structure of the original one: nodes that play a role of being structurally similar to many others need not have specific requirements concerning its degree. Regarding eigenvector centrality, we can observe a non-linear tendency between both networks. The interpretation of the highest scores of eigenvector centrality in the dual network is the following: nodes that are, not only structurally similar to many other nodes, but whose neighbours are so (and the neighbours of their neighbours, and so on). Conversely, the nodes with low values of eigenvector centrality are those which are unique and which neighbours so are. Note that the values of the centralities in both the original and the dual network are positively correlated up to a threshold, from which a slightly negative correlation appears.

Despite the ranking of node importance obtained from centralities provides us with the more relevant information about structural similarity, the distribution of scores is rather homogeneous. In order to obtain a more clear pattern, we suggest using the binnarized dual network, $\mathcal{H}_{\mathcal{B}}$, in order to compute network measures, such as centralities or community detection, because we get rid of non-significant low-weighted edges and the network becomes more sparse, a characteristic which leads to a better separation of the roles of nodes. On this basis, Figs \ref{fig:macaque_Degree_Sparse} and \ref{fig:macaque_Eigenvector_Sparse} show the results of degree and eigenvector centralities using $\mathcal{H}_{\mathcal{B}}$, the binnarized dual network. In this case, the ranking of nodes is similar to that of the weighted dual network but differences between nodes are more emphasized.

The brain regions with lower eigenvector centrality on the dual network are the amygdala, inferior temporal cortex, inferior temporal cortex, primary visual cortex, anterior visual area, ventral part and hippocampus. The regions with the highest eigenvector centrality score are the prefrontal polar cortex, primary auditory cortex, posterior cingulate cortex, posterior insula, orbitolateral prefrontal cortex and superior parietal cortex.
\subsection{Quasi-symmetric communities.}
\label{subsec:Quasi-SymmetriesDetection}
The classification of nodes into perfectly symmetric clusters or orbits has been already addressed and solved in the field of discrete algebra. In Section \ref{subsec:SymmetriesDetection} we suggest an alternative approach to obtain these same results based on a dynamic model. Nonetheless, we are interested in classifying nodes into different communities based on the structural similarity, and not perfect equivalence, between them. This problem is a particular case of the more general topic of unsupervised classification algorithms, where no correctly classified samples are provided. In other words, we do not know the number of groups and the characteristics of the nodes belonging to each group. However and differently to classical classification problems, our main point is relying on the detection of the number of peaks of the Gaussian Kernel Estimator distribution fitted on the scaled phases (see Section \ref{subsubsec:countingQuasiSymmetries}). For large enough networks (those which considering a distribution makes sense), the number of peaks will be considered as the number of expected communities that the community detection algorithm should obtain. Hence, only the classification of nodes in each community is missing. To address this question, several approaches are proposed, although we do not reject alternative methodologies that may come up.
\begin{itemize}
    \item Distance based approach: in section \ref{subsubsec:degree} we explore the distribution of phases by fitting a Kernel density distribution in order to decide whether the structural similarity of a network has a richer pattern than a random network. Once the optimal bandwidth of the Gaussian kernel is numerically computed, we can easily count the number of peaks of the distribution (see, for example, Fig. \ref{fig:12symmetriesDistribution}(b)). Our method consists in considering this last value as the number of expected communities in the corresponding parameter of an unsupervised clustering algorithm, for example, k-means clustering or hierarchical clustering (following Reference \cite{Nicosia2013a}), and obtain the optimal partition into communities. Note that the algorithm delivers the cluster to which each node belongs to, but not in a network-like structure. 
    
Figure \ref{fig:5symmetriesKmeans}(a,c) shows the dual network corresponding to the networks with originally $5$ symmetries with no perturbation and $w_{max}=0.05$, respectively. The position of nodes are computed using the Fruchterman-Reingold force-directed algorithm \cite{Fruchterman1991} considering $\mathcal{H}$. Colors represent the distinct clusters obtained using k-means algorithm with the number of clusters given by the number of peaks of the Kernel density distribution, i.e., 5. (see the upper panel in  Fig. \ref{fig:5symmetriesDistribution}(b)). Similarly, Fig. \ref{fig:12symmetriesKmeans}(a,c) shows the obtained communities for the networks with $12$ perfect symmetries and after applying a random noise with $w_{max}=0.05$, respectively. In order to verify whether all nodes are correctly classified into the different clusters (a number which is given by the detected peaks of the KDE distributions), we plot the obtained scaled phases of each node and its corresponding membership into the different communities. For the case of no perturbation, single points are expected, as nodes belonging to the same cluster collapse into a single phase value (see Figs \ref{fig:5symmetriesKmeans}(b) and \ref{fig:12symmetriesKmeans}(b)). For perturbed networks, a dispersion of phases around different centroids is expected (see Figs \ref{fig:5symmetriesKmeans}(d) and \ref{fig:12symmetriesKmeans}(d)).
\begin{figure}[H]
\centering
\includegraphics[width=\textwidth]{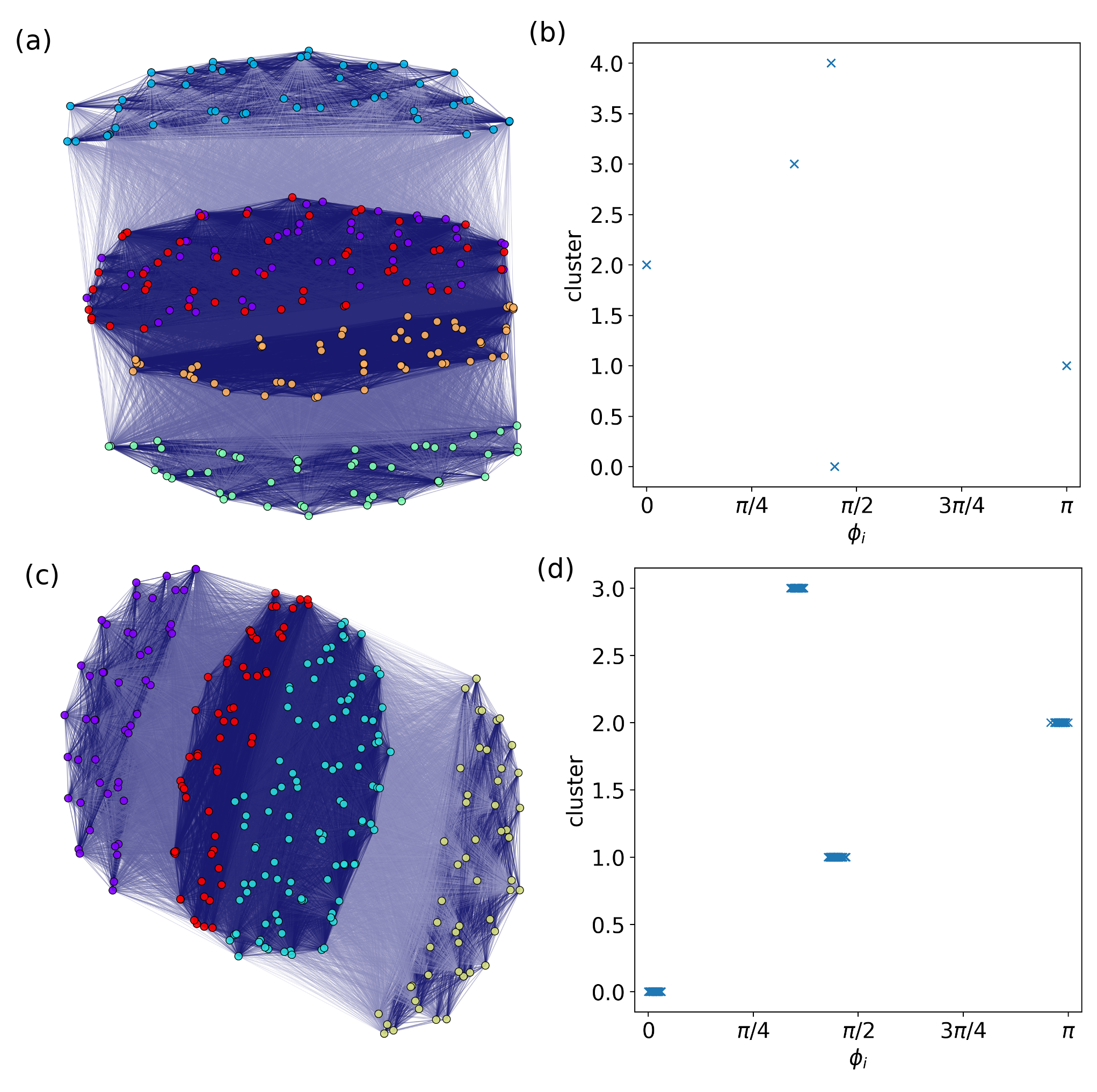}
\caption{(a) Dual network corresponding to the network (264 nodes) with originally 5 symmetries and no perturbation applied ($w_{max}=0.0$). Colors represent the distinct clusters obtained using k-means algorithm with the number of clusters given by the number of peaks of the Kernel density distribution, i.e., 5. (see the upper panel in  Fig. \ref{fig:5symmetriesDistribution}(b)). (b) Corresponding scatter plot of the scaled phases versus the cluster ID. Nodes are correctly classified, as the number of distinct groups is 5, as expected for the case of a network with 5 symmetries and no perturbation applied. (c) Dual network obtained after applying a perturbation of $w_{max}=0.05$ to the original network. The number of clusters is 4  (see the third panel Fig. \ref{fig:5symmetriesDistribution}(b)). (d) Corresponding scatter plot of the scaled phases versus the cluster ID. Notice that phases corresponding to nodes that belong to the same cluster have a dispersion different than zero.}
\label{fig:5symmetriesKmeans}
\end{figure}
\begin{figure}[H]
\centering
\includegraphics[width=\textwidth]{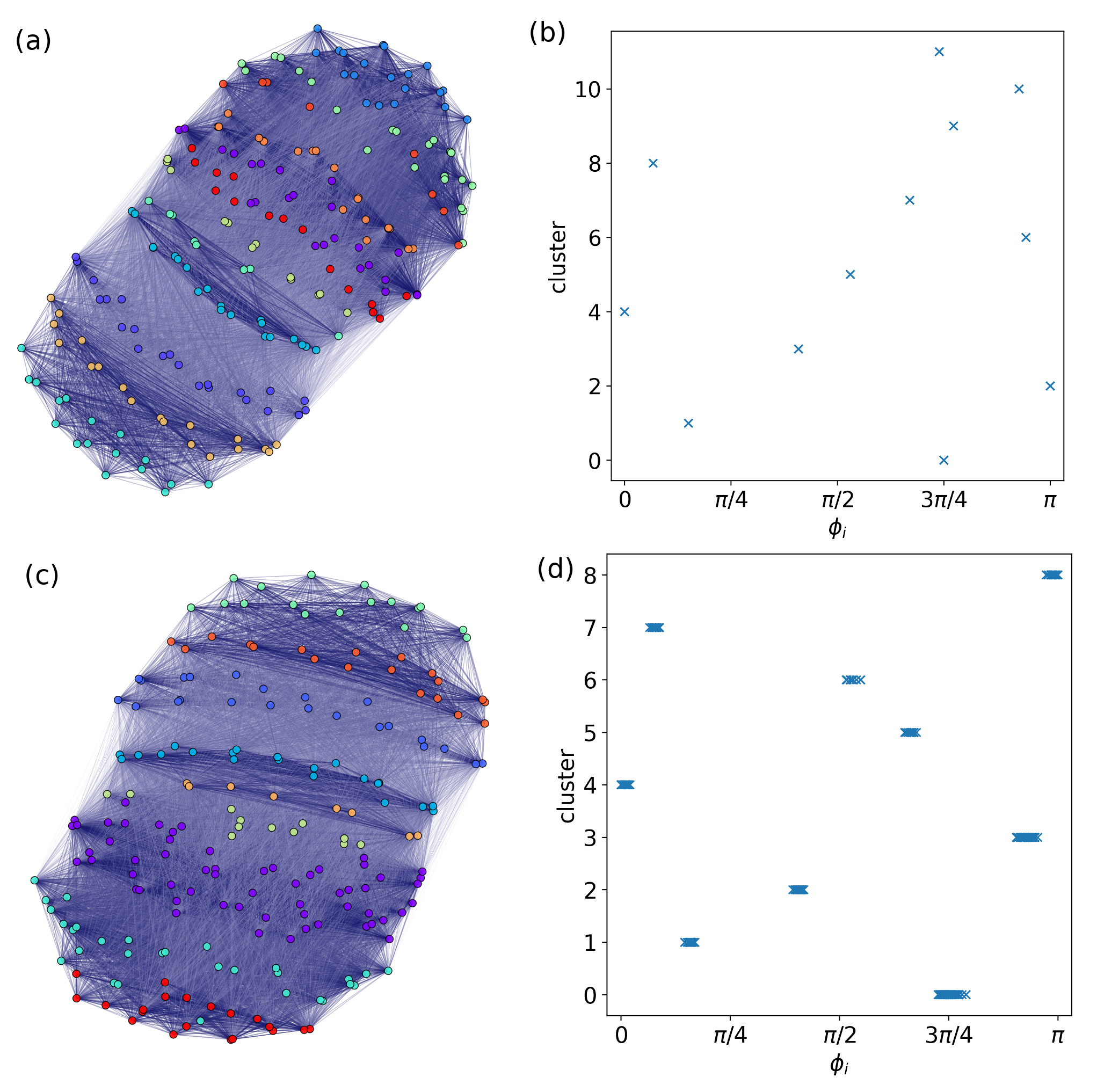}
\caption{(a) Dual network corresponding to the network (209 nodes) with originally 12 symmetries and no perturbation applied ($w_{max}=0.0$). Colors represent the distinct clusters obtained using k-means algorithm with the number of clusters given by the number of peaks of the Kernel density distribution, i.e., 12. (see the upper panel in  Fig. \ref{fig:12symmetriesDistribution}(b)). (b) Corresponding scatter plot of the scaled phases versus the cluster ID. Nodes are correctly classified, as the number of distinct groups is 12, as expected for the case of a network with 12 symmetries and no perturbation applied. (c) Dual network obtained after applying a perturbation of $w_{max}=0.05$ to the original network. The number of clusters is 9  (see the third panel Fig. \ref{fig:12symmetriesDistribution}(b)). (d) Corresponding scatter plot of the scaled phases versus the cluster ID. Notice that phases corresponding to nodes that belong to the same cluster have a dispersion different than zero.}
\label{fig:12symmetriesKmeans}
\end{figure}

Alternatively to k-means clustering algorithm we could apply hierarchical clustering in order to split the nodes into communities and obtain equivalent results. 

The classification of items into clusters obtained by applying unsupervised algorithms, such as k-means or hierarchical clustering, depend on the choice of the number of clusters. The problem of the optimal choice has been widely studied and several approaches have been proposed in order to select the proper number of clusters or the cut height, for the cases of k-means and hierarchical clustering, respectively. We compare the results of the number of peaks obtained by using the Kernel density distribution approach with that obtained by choosing the optimal number of clusters with the most common method: the elbow curve \cite{Ketchen1996}. The obtained optimal number of clusters does not coincide with our approach as, when the network is not perturbed, larger number of clusters are suggested. The Kernel density distribution approach automatically detects the optimal bandwidth and adapts to each distribution, from narrow peaks to broad and diffuse distributions. 
    \item Dual Network approach:  using the definition of the binnarized dual network, in Eq.(\ref{eq:binnaryDual}), we choose $w_{threshold}$ such that the number of detected communities using the Louvain algorithm equals the number of detected peaks. In Figs \ref{fig:5symmetries_HB} and \ref{fig:12symmetries_HB}, an example is provided for the networks with originally $5$ and $12$ symmetries, respectively.
    
    One key benefit of using $\mathcal{H}_{\mathcal{B}}$ is working with a sparse network and keeping only significant relations between similar nodes. On this basis, in Fig. \ref{fig:macaque_communities} we present the result of applying community detection on the binnarized dual network of the Macaque brain network, with the positions of the layout being determined by the original network. Notice that the left-right hemispheres separation is recovered from the communities and similar regions are gathered in the same \textit{quasi-symmetric} community. Nodes belonging to the same community play a similar role or have a similar structural pattern.
    
\end{itemize}
\section{Discussion}
There have been a number of attempts to deal with approximate symmetries in networks. Beyond structural or topological symmetry, one should consider the fact that real-world networks are exposed to fluctuations or errors, as well as mistaken insertions or removals. Understanding network (approximate) symmetry is of great relevance for the analysis of real-world networks, as they have a significant effect on network dynamics and function. In the present work, we provide an alternative notion to approximate symmetries, which we call `quasi-symmetries'. Differently from other definitions, quasi-symmetries remain free to impose any invariance of a particular network property and are obtained from an oscillatory dynamical model: the Kuramoto-Sakaguchi model. 

A first main contribution is exploring the distributions of structural similarity among all pairs of nodes and finding a benchmark to determine whether a network has a more complex pattern to that of a random network concerning quasi-symmetries: the criteria consists in determining whether the number of quasi-symmetric groups is greater than one. The number of peaks is derived from the Gaussian Kernel Density Estimator (KDE) (a detailed explanation of the KDE is found in the appendix \ref{ap:KDE}). Despite we have used this approach, we are open to alternative methodologies to find a more accurate detection of the number of peaks. Moreover, other Kernels may be considered. 

Secondly, we define the `dual network', a weighted network (and its corresponding binnarized counterpart) that effectively encloses all the information of quasi-symmetries in the original one. The dual network allows for the analysis of centrality measures and community detection. The first informs us about the nodes that play a unique role in the network and of those that behave similarly to many other nodes. The latter leads to a classification of nodes into quasi-symmetric communities, the natural extension of the automorphism group orbits (structurally symmetric nodes) of a network. The use of the binary dual network, $\mathcal{H}_{\mathcal{B}}$, is advantageous as it leads to more heterogeneous results in the ranking of node importance and it enables a more significant classification of nodes into quasi-symmetric communities. The number of detected peaks in the KDE distribution determines the family of feasible $\mathcal{H}_{\mathcal{B}}$. However, one could suggest other criteria, as well as threshold models, in order to create the binnarized counterpart of the dual network.

Finally, we state that in the present work we bring out a general framework to deal with approximate symmetries in complex networks. The dual network has been presented as a useful tool to work with quasi-symmetries and a number of applications have been addressed. Nevertheless, there is a lot of room for obtaining other interesting insights. The analysis of network tolerance to attack to the quasi-symmetric structure or the analysis of quasi-symmetries in multilayer networks are some examples.

\label{Conclusions}
\section*{Funding}
The authors acknowledge financial support from
MINECO via Project No.PGC2018-094754-B-C22
(MINECO/FEDER,UE) and Generalitat de Catalunya
via Grant No. 2017SGR341. G.R.-T. also acknowledges
MECD for Grant No. FPU15/03053 and Universitat de
Barcelona for grant PREDOCS-UB 2019.
\section*{Acknowledgements}
The final version of this work is published on the \textit{Journal of Complex Networks}, Volume 9, Issue 3, June 2021, https://doi-org.sire.ub.edu/10.1093/comnet/cnab025.
\appendix
\section{Steady-state solution of the linearized Kuramoto-Sakaguchi model}
\label{ap:KMLinear}
Consider the dynamics described by the Kuramoto-Sakaguchi model as defined in Eq.(\ref{eq:KSmodel}). If the system reaches a synchronized state, that is, $\dot{\theta_i} = \omega \ \forall i$, and $\alpha$ is small enough, Eq.(\ref{eq:KSmodel}) can be linearized as follows
\begin{eqnarray*}
    \dot{\theta}_i\sim\omega+K\sum_j A_{ij}(\theta_j-\theta_i-\alpha) = \nonumber \\ = \omega+K\sum_j A_{ij}\theta_j -K\sum_jd_i\theta_j-K\alpha d_i
\end{eqnarray*}where we have used $\sum_j A_{ij}=d_i$ and $d_i$ is the degree of the $i$th node. Using the definition of the Laplacian matrix $L_{ij}=d_{i}\delta_{ij}-A_{ij}$
\begin{equation}
\label{eqap:KSmodelLinear}
    \dot{\theta}_i\sim\omega-K\sum_j L_{ij}\theta_j -K\alpha d_i, \ j\in \Gamma_{i}
\end{equation}
In the steady state under the linear approximation, a common frequency, $ \dot{\theta}_i=\Omega$, is achieved for all oscillators. Summing over $i$ index we obtain
\begin{equation*}
 \sum_i\Omega = \sum_i\omega-K\sum_i\sum_j L_{ij}\theta_j -K\sum_i\alpha d_i
\end{equation*} and,
\begin{equation*}
 N\Omega = N\omega-K\sum_i\alpha d_i
\end{equation*}where we have applied the row-sum equal to zero property of the Laplacian matrix $\sum_i L{ij}=0$. Finally,
\begin{equation}
    \label{eqap:omega}
    \Omega = \omega - K\alpha \left\langle d \right\rangle
\end{equation}where $\left\langle d \right\rangle \equiv \sum_i d_i/N$. Substituting Eq.(\ref{eqap:omega}) into Eq.(\ref{eqap:KSmodelLinear}) we obtain the values of the phases at any time
\begin{equation}
    \sum_jL_{ij}\theta_j = \alpha (\left\langle d \right\rangle-d_i)
\end{equation}or, in matrix notation,
\begin{equation}
\label{eqap:phasesMatrixNotation}
    L \vec{\theta} = \alpha (\left\langle d \right\rangle \vec{\mathrm{1}}-\vec{d})
\end{equation}where $\vec{[d]}_i = d_i$.
\section{Condition on a diagonal matrix so that it commutes with an automorphism permutation matrix}
\label{ap:commute}
We provide a proof of the condition that a diagonal matrix must meet in order to commute with the permutation matrix, $P$, corresponding to an automorphism $\sigma \in Aut(\mathcal{G})$. This last statement is needed to proof that the Laplacian matrix of a network also commutes with the permutation matrix, $P$:
\begin{proof}
\label{proof:LaplacianCommutes}
Let $P$ be a permutation matrix corresponding to an automorphism, $\sigma \in Aut(\mathcal{G})$. Hence,
\begin{equation}
\label{eqap:canonicalBase}
    Pe_i = e_{\sigma(i)} \ \forall i \in \{0,...,n-1 \}
\end{equation}where $\{e_0,...,e_{n-1}\}$ denotes the standard basis of $\mathbf{R}^n$. For a $n\times n$ matrix $M$ to commute with $P$ we must have $MP=PM$, or equivalently, $P^{-1}MP=M$. If $M$ is the diagonal matrix 
\begin{equation*}
    M =
  \begin{pmatrix}
    m_{0} & & \\
    & \ddots & \\
    & & m_{n-1}
  \end{pmatrix}
\end{equation*}for all $i \in \{0,...,n-1 \}$, 
\begin{eqnarray}
P^{-1}MPe_i=P^{-1}Me_{\sigma(i)}=P^{-1} m_{\sigma(i)}e_{\sigma(i)} = \nonumber\\
=  m_{\sigma(i)}P^{-1}e_{\sigma(i)} = m_{\sigma(i)}e_i
\end{eqnarray} where we have used $e_i = P^{-1}e_{\sigma(i)}$ derived from Eq.(\ref{eqap:canonicalBase}) and the fact that $m_{\sigma(i)}$ is a scalar.
From this we can write
\begin{equation}
\label{eqap:conditionCommute}
    P^{-1} M P = M_{\sigma}
\end{equation}where 
\begin{equation*}
    M_{\sigma} \equiv
  \begin{pmatrix}
    m_{\sigma(0)} & & \\
    & \ddots & \\
    & & m_{\sigma(n-1)}
  \end{pmatrix}
\end{equation*}
So $M=diag(m_0,...,m_{n-1})$ commutes with $P$ if and only if $m_{\sigma(i)}=m_i$ for all $i$. But the condition in Eq.(\ref{eqap:conditionCommute}) holds as long as $m_i=m_j$ for all $i,j$ that belong to the same orbit.
\end{proof}
 
\section{The bi-conditional proof of the statement `nodes with equal $\phi$ belong to the same orbit'}
\label{ap:biconditional}
We derive the required conditions for the statement `Nodes that have the same phases belong to the same orbit' to be true. As we will see, the implication of two nodes having the same phases is, in most cases, that these nodes belong to the same orbit. However, there might be some cases where the equality of phases is achieved by other conditions.

Suppose nodes $i$ and $j$ have the same phase, $\theta_i = \theta_j$ or $\phi_i = \phi_j$. Then, from Eq.(\ref{eq:phasesReduced}) we can write the corresponding solutions using the reduced Laplacian
\begin{equation}
\phi_i = \alpha \sum_k [\tilde{L}^{-1}]_{ik} (\left\langle d \right\rangle - d_i) \text{ and } \phi_j = \alpha \sum_k [\tilde{L}^{-1}]_{jk} (\left\langle d \right\rangle - d_j)
\end{equation}
The condition of identical phases leads to the equality
\begin{equation}
\sum_k [\tilde{L}^{-1}]_{ik} (\left\langle d \right\rangle - d_i) =  \sum_k [\tilde{L}^{-1}]_{jk} (\left\langle d \right\rangle - d_j) \Rightarrow (\left\langle d \right\rangle - d_i) \sum_k [\tilde{L}^{-1}]_{ik}  =   (\left\langle d \right\rangle - d_j)\sum_k [\tilde{L}^{-1}]_{jk} 
\label{eq:proof_1}
\end{equation}because $d_i$ and $d_j$ do not depend on $k$.

Condition (\ref{eq:proof_1}) comes from assuming $\phi_i=\phi_j$.

We consider two different cases: the considered nodes having the same degree or different degree.

\begin{itemize}
\item $d_i = d_j$

If nodes $i$ and $j$ have the same degree, from Eq.(\ref{eq:proof_1}) we get $\sum_k [\tilde{L}^{-1}]_{ik}  = \sum_k [\tilde{L}^{-1}]_{jk} $. This last equality is only true when $i$ and $j$ belong to the same orbit. 

Therefore, the straightforward implication is that $i$ and $j$ having the same phase implies that $i$ and $j$ belong to the same orbit.

\item $d_i \neq d_j$

This case is more tricky. We can write the relations between the sum along rows of the inverse of the reduced Laplacian as
\begin{equation}
\frac{\sum_k [\tilde{L}^{-1}]_{ik}}{\sum_k [\tilde{L}^{-1}]_{jk}} = \frac{\left\langle d \right\rangle - d_i}{\left\langle d \right\rangle - d_j}
\label{eq:proof_2}
\end{equation}
Making use of the inequality $\sum_k [\tilde{L}^{-1}]_{ik} > 0 \ \forall i$, we get $\displaystyle\frac{\sum_k [\tilde{L}^{-1}]_{ik}}{\sum_k [\tilde{L}^{-1}]_{jk}} \geq 0$. Finally,
\begin{equation}
\frac{\left\langle d \right\rangle - d_i}{\left\langle d \right\rangle - d_j} \geq 0
\label{eq:proof_3}
\end{equation}
Considering that $d_i > 0 \ \forall i$, the inequality (\ref{eq:proof_3}) has the following solution:
\begin{equation}
d_i \geq \left\langle d \right\rangle \text{ and } d_j \geq \left\langle d \right\rangle \text{ or } 0 < d_i \leq \left\langle d \right\rangle \text{ and } 0 < d_j \leq \left\langle d \right\rangle 
\label{eq:proof_4}
\end{equation} where we have simplified by considering that the network is large enough.

From this second case we can conclude that the equality (\ref{eq:proof_1}) can be achieved when $d_i\neq d_j$ only if Eq.(\ref{eq:proof_2}) and Eq.(\ref{eq:proof_4}) are true. These last requirements represent very strong restrictions for a network. Firstly, the fine tuning (only feasible for weighted networks) of the degree sequence implied in Eq.(\ref{eq:proof_2}) is hard to be achieved. Secondly, the inequality (\ref{eq:proof_4}) adds further constrains on the first condition.
\end{itemize}

To conclude, we can say that the double implication `Nodes that have the same phases $ \Longleftrightarrow$  Nodes that belong to the same orbit' is always true except by the cases where a pair of nodes $i$ and $j$ that have different degrees, $d_i \neq d_j$ meet the conditions expressed in Eqs.(\ref{eq:proof_2}) and (\ref{eq:proof_4}). Note that restriction (\ref{eq:proof_2}) requires that a quotient of degrees takes a specific value, resulting from a  non-linear transformation of network parameters, and hence, it is highly unlikely. From a probabilistic perspective, the probability that a continuous random variable takes a specific value is zero.
\color{black}
\section{Kernel Density Estimator}
\label{ap:KDE}
Kernel Density Estimator (KDE) is a non-parametric standard technique in explorative data analysis to estimate the probability density function of a random variable first introduced in References  \cite{Rosenblatt1956} and \cite{Parzen1962}. From a finite data sample the probability function of the whole population is infered. The KDE method takes a kernel and a parameter, the bandwidth, that affects the level of smoothness the resulting curve has.

The problem can be posed as `How can one estimate a probability density function $f(x)$ given a sequence of $n$ independent identically distributed random variables $X_1,..., X_n$ from this density $f$? \cite{Turlach1993} The estimator of $f$, $\hat{f}_h(x)$ is defined by
\begin{equation}
\label{KDE}
\hat{f}_h(x) = \frac{1}{n}\sum_{i=1}^{n}K_h(x-X_i)
\end{equation}where $h$ is the smoothing parameter called the bandwidth and $K$ is the kernel, $K_h(u)=K(u/h)/h$. The Kernel is imposed to be symmetric and non-negative, and $K$ itself being a probability density. Then, $\hat{f}_h(x)$ intuitively places at each observation point $X_i$ a probability mass according to $K_h$ and then averages. Some of the commonly kernels are uniform, triangle,quartic, triweight, Epanechnikov and Gaussian. It turns out that the choice of the bandwidth is much more important for the estimation of the density than the particular shape of the kernel. Small values of $h$ result into an over-fitted density distribution, showing spurious features ot the latter, while to big values of $h$ lead to an estimate which is too biased and may not reveal relevant features of the distribution.

The construction of a kernel density estimate finds and  interpretation in thermodynamics, since the Gaussian KDE is the solution of the heat propagation model, i.e., the solution of the estimator is equivalent to the amount of heat generated when heat kernels are placed at each data point locations \cite{Botev2010}. Note that the Gaussian kernel density estimator is the unique solution to the diffusion partial differential equation PDE
\begin{equation}
\label{eq:diffusionPDE}
\frac{\partial}{\partial t} \hat{f}(x,t) = \frac{1}{2} \frac{\partial ^2}{\partial^2 x} \hat{f}(x,t), \ t>0
\end{equation} with initial condition $\hat{f}(x,0)=1/N \sum_i \delta(x-X_i)$ is the empirical density of the data and $t = \sqrt{h}$.
Eq.(\ref{eq:diffusionPDE}) corresponds to the Fourier heat equation.

We however make a few comments on some drawbacks of the popular Gaussian Kernel density estimator: firstly, it lacks local adaptativity, which often leads to substantial sensitivity to outliers as well as a tendency to flatten the peaks and valleys of the distribution \cite{Terrell1992}. Secondly, just as most kernel estimators, it suffers from boundary bias, as most kernels do not take into account further information about the domain of the data, i.e., data being nonnegative. Some of these issues have been addressed by introducing more complex higher-order kernels.

In this work, we pursue to hold the methodology as simple as possible. To this end and because it meets the objectives of the posed problem, we use the Gaussian kernel. Nonetheless, we suggest the reader to consider implementing a kernel based on diffussive processes, in Reference \cite{Botev2010}, as it solves the mentioned problems of standard kernels estimators.
\section{Whole-cortex Macaque structural connectome: results}
\label{ap:macaqueResults}
In several section we have applied our measures to the whole-cortex macaque structural connectome constructed from a combination of axonal tract-tracing and diffusion-weighted imaging data \cite{macaque2018}. We present the corresponding figures of the results concerning centralities and community detection.
\begin{figure}[H]
    \centering
    \includegraphics[width=\textwidth]{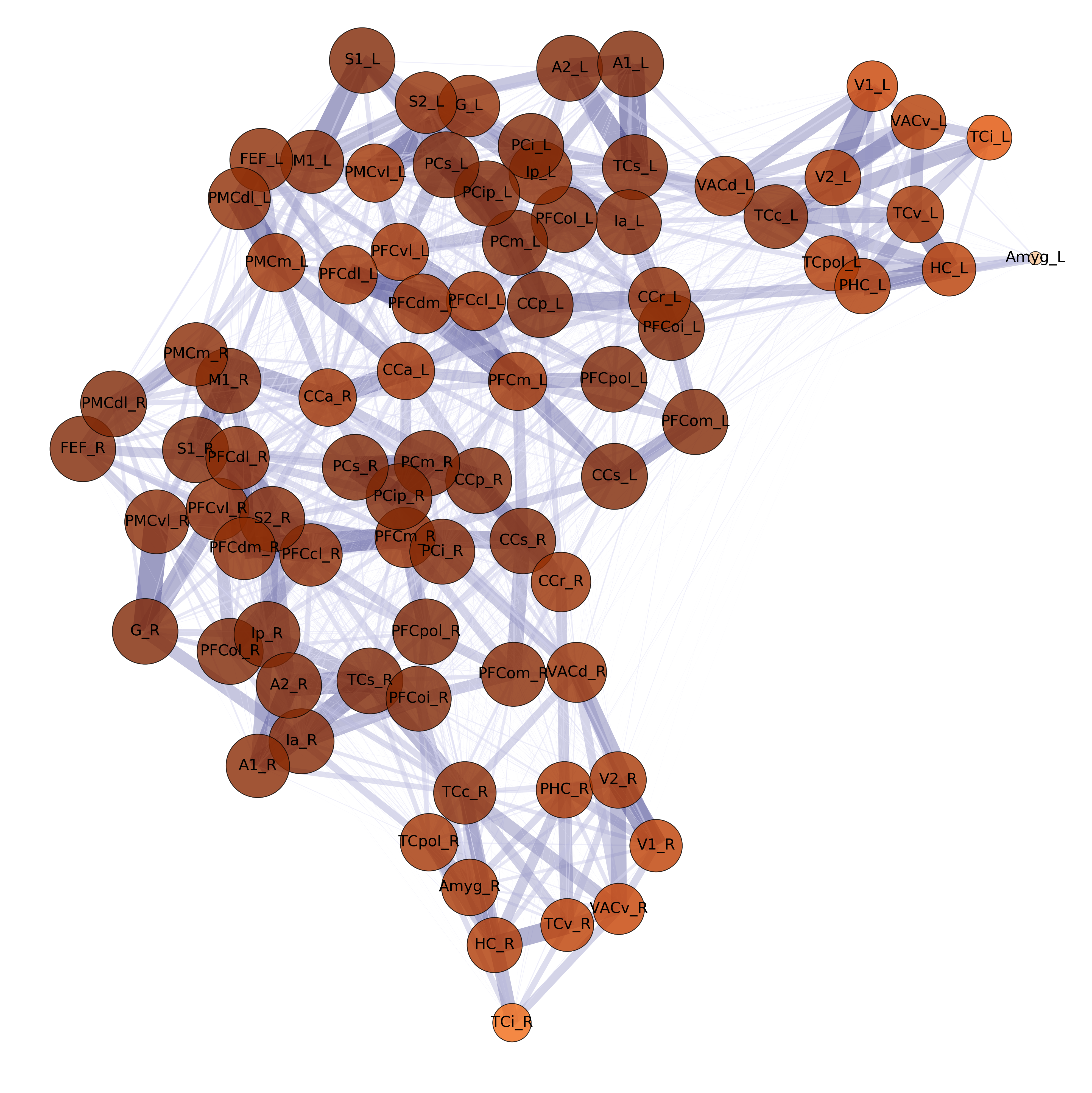}
    \caption{$\mathcal{H}$ network of the Macaque brain network, such that the radius of the nodes is proportional to the values of the degree centrality of the dual network, and the color map is built such that darker values correspond to larger values of this centrality. The position of nodes are computed using the Fruchterman-Reingold force-directed algorithm \cite{Fruchterman1991} considering the original network.}
    \label{fig:macaque_Degree_NonSparse}
\end{figure}
\begin{figure}[H]
    \centering
    \includegraphics[width=\textwidth]{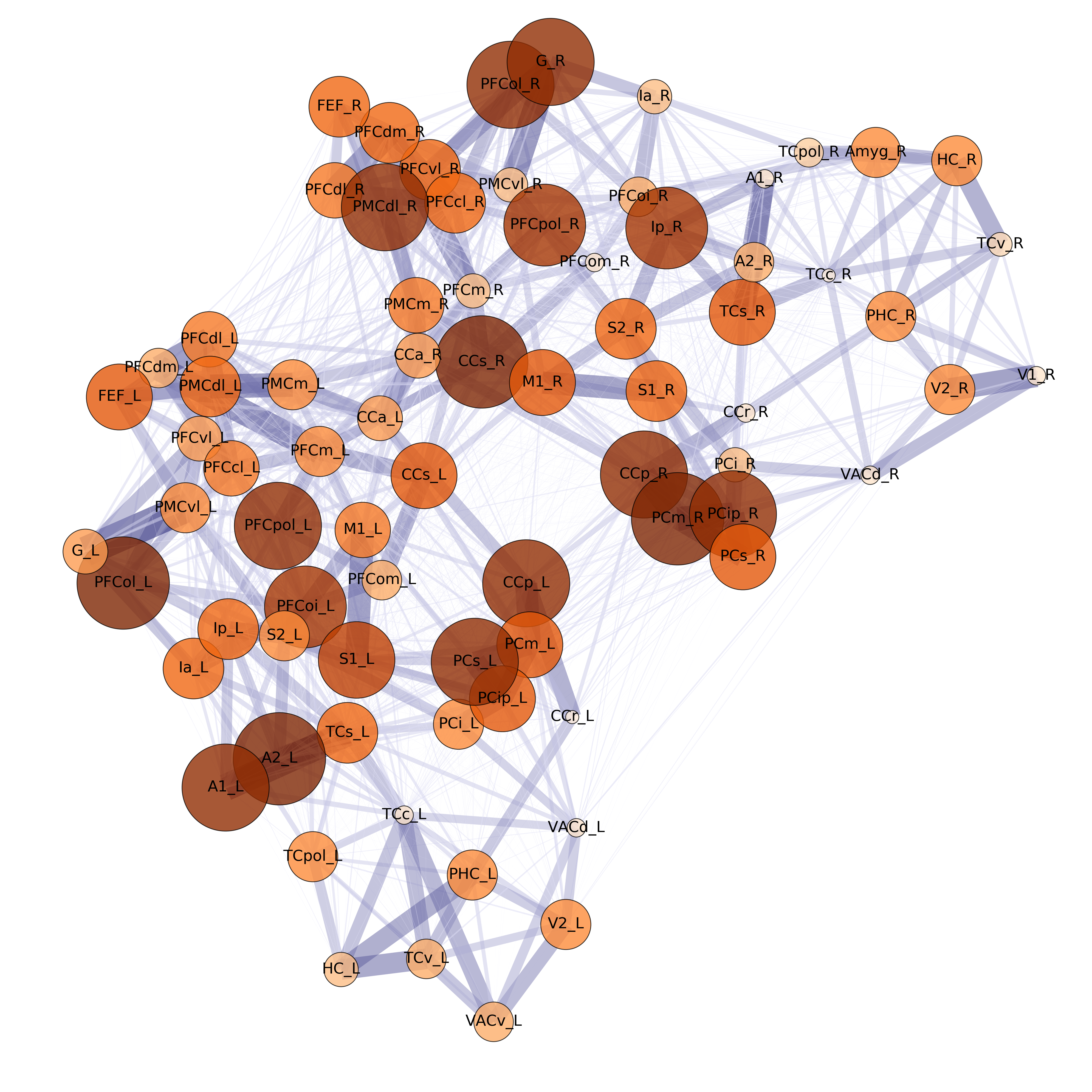}
    \caption{$\mathcal{H}_{\mathcal{B}}$ network of the Macaque brain network, such that the radius of the nodes is proportional to the values of the degree centrality of the dual network, and the color map is built such that darker values correspond to larger values of this centrality. The position of nodes are computed using the Fruchterman-Reingold force-directed algorithm \cite{Fruchterman1991} considering the original network. The nodes above the 95\% percentile of phases are removed (outliers).}
    \label{fig:macaque_Degree_Sparse}
\end{figure}
\begin{figure}[H]
    \centering
    \includegraphics[width=\textwidth]{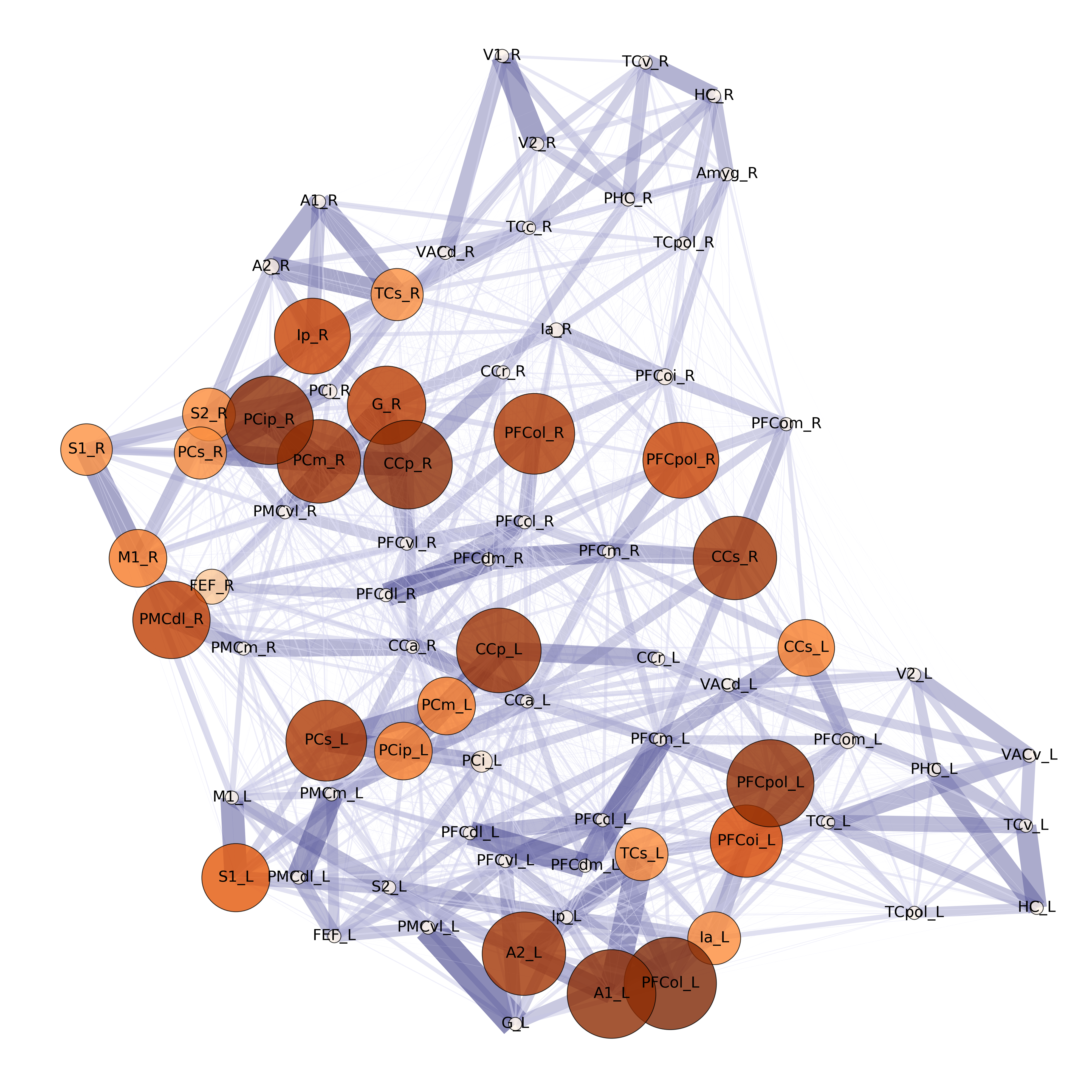}
    \caption{$\mathcal{H}_{\mathcal{B}}$ network of the Macaque brain network, such that the radius of the nodes is proportional to the values of the eigenvector centrality of the dual network, and the color map is built such that darker values correspond to larger values of this centrality. The position of nodes are computed using the Fruchterman-Reingold force-directed algorithm \cite{Fruchterman1991} considering the original network. The nodes above the 95\% percentile of phases are removed (outliers).}
    \label{fig:macaque_Eigenvector_Sparse}
\end{figure}
\begin{figure}[H]
        \centering
        \includegraphics[width=\textwidth]{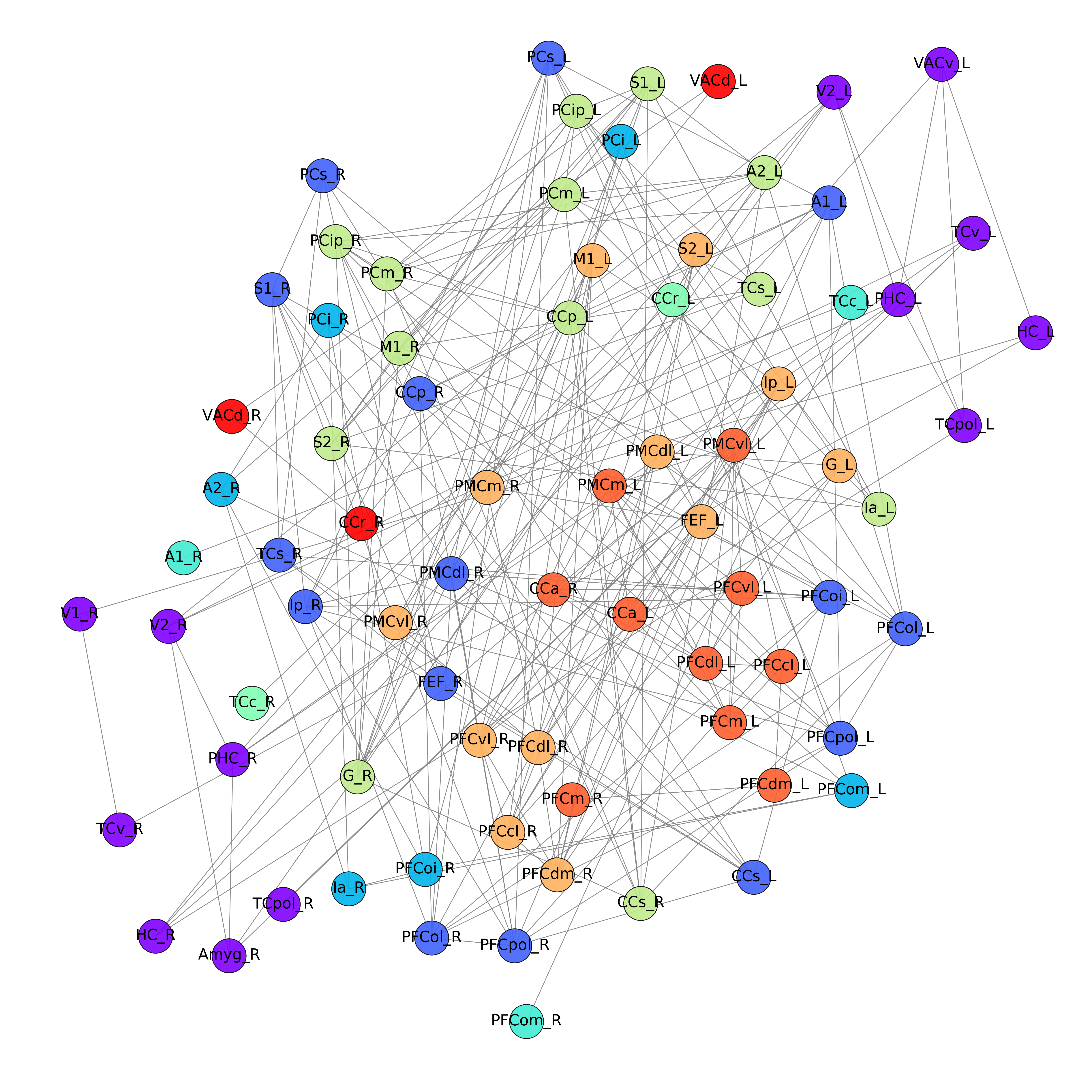}
        \caption{$\mathcal{H}_{\mathcal{B}}$ network of the macaque brain network with $w_{threshold}=0.97$. Colors correspond to different communities. The position of nodes are computed using the Fruchterman-Reingold force-directed algorithm \cite{Fruchterman1991} considering the original network.}
        \label{fig:macaque_communities}
    \end{figure}

\bibliographystyle{ieeetr}

\begin{thebibliography}{99}

\bibitem{Caldarelli2007}
G.~Caldarelli, {\em Scale-Free Networks: Complex Webs in Nature and
  Technology}.
\newblock Oxford University Press, May 2007.

\bibitem{Barabasi2009}
A.-L. Barabasi, ``{Scale-Free Networks: A Decade and Beyond},'' {\em Science},
  vol.~325, no.~5939, pp.~412--413, 2009.

\bibitem{Caldarelli2007a}
G.~Caldarelli and A.~Vespignani, {\em Large Scale Structure and Dynamics of
  Complex Networks}.
\newblock {WORLD} {SCIENTIFIC}, June 2007.

\bibitem{PastorSatorras2004}
R.~Pastor-Satorras and A.~Vespignani, {\em Evolution and Structure of the
  Internet}.
\newblock Cambridge University Press, Feb. 2004.

\bibitem{Newman2001}
M.~E.~J. Newman, ``Scientific collaboration networks i: Network construction
  and fundamental results,'' {\em Physical Review E}, vol.~64, June 2001.

\bibitem{Watts1998b}
D.~J. Watts and S.~H. Strogatz, ``{Collective dynamics of ‘small-world'
  networks},'' {\em Nature}, vol.~393, pp.~440--442, jun 1998.

\bibitem{Albert2002}
R.~Albert and A.-L. Barab{\'{a}}si, ``Statistical mechanics of complex
  networks,'' {\em Reviews of Modern Physics}, vol.~74, pp.~47--97, Jan. 2002.

\bibitem{Boccaletti2006}
S.~Boccaletti, V.~Latora, Y.~Moreno, M.~Chavez, and D.-U. Hwang, ``{Complex
  networks: Structure and dynamics},'' {\em Physics Reports}, vol.~424,
  pp.~175--308, 2006.

\bibitem{Strogatz2001}
S.~H. Strogatz, ``{Exploring complex networks},'' {\em Nature}, vol.~410,
  pp.~268--276, mar 2001.

\bibitem{Girvan}
M.~Girvan and M.~E.~J. Newman, ``{Community structure in social and biological
  networks},'' {\em Proceedings of the National Academy of Sciences}, vol.~99,
  no.~12, pp.~7821--7826, 2002.

\bibitem{Smith2019}
D.~Smith and B.~Webb, ``Hidden symmetries in real and theoretical networks,''
  {\em Physica A: Statistical Mechanics and its Applications}, vol.~514,
  pp.~855--867, Jan. 2019.

\bibitem{Sanchez-Garcia2020}
R.~J. S{\'{a}}nchez-Garc{\'{i}}a, ``{Exploiting symmetry in network
  analysis},'' {\em Communications Physics}, vol.~3, no.~1, 2020.

\bibitem{Liu2011}
Y.-Y. Liu, J.-J. Slotine, and A.-L. Barab{\'{a}}si, ``Controllability of
  complex networks,'' {\em Nature}, vol.~473, pp.~167--173, May 2011.

\bibitem{Liu2013}
Y.-Y. Liu, J.-J. Slotine, and A.-L. Barabasi, ``Observability of complex
  systems,'' {\em Proceedings of the National Academy of Sciences}, vol.~110,
  pp.~2460--2465, Jan. 2013.

\bibitem{Whalen2015}
A.~J. Whalen, S.~N. Brennan, T.~D. Sauer, and S.~J. Schiff, ``Observability and
  controllability of nonlinear networks: The role of symmetry,'' {\em Physical
  Review X}, vol.~5, Jan. 2015.

\bibitem{Nicosia2013a}
V.~Nicosia, M.~Valencia, M.~Chavez, A.~D{\'{i}}az-Guilera, and V.~Latora,
  ``{Remote Synchronization Reveals Network Symmetries and Functional
  Modules},'' {\em Physical Review Letters}, vol.~110, no.~174102, 2013.

\bibitem{Pecora2014}
L.~M. Pecora, F.~Sorrentino, A.~M. Hagerstrom, T.~E. Murphy, and R.~Roy,
  ``Cluster synchronization and isolated desynchronization in complex networks
  with symmetries,'' {\em Nature Communications}, vol.~5, June 2014.

\bibitem{Jiang2016}
X.~Jiang and D.~M. Abrams, ``{Symmetry-broken states on networks of coupled
  oscillators},'' {\em Physical Review E}, vol.~93, no.~5, pp.~1--5, 2016.

\bibitem{Zee1986-ZEEFST}
A.~Zee, {\em Fearful Symmetry: The Search for Beauty in Modern Physics}.
\newblock Princeton University Press, 1986.

\bibitem{Lockwood1978}
E.~H. Lockwood and R.~H. Macmillan, {\em Geometric symmetry}.
\newblock Cambridge University Press Cambridge ; New York, 1978.

\bibitem{Garlaschelli2010}
D.~Garlaschelli, F.~Ruzzenenti, and R.~Basosi, ``Complex networks and symmetry
  i: A review,'' {\em Symmetry}, vol.~2, pp.~1683--1709, Sept. 2010.

\bibitem{Stewart2004}
I.~Stewart, ``Networking opportunity,'' {\em Nature}, vol.~427, pp.~601--604,
  Feb. 2004.

\bibitem{Olver2016}
P.~Olver, ``The symmetry groupoid and weighted signature of a geometric
  object,'' {\em Journal of Lie Theory}, vol.~26, pp.~235--267, Jan. 2016.

\bibitem{Holme2006}
P.~Holme, ``Detecting degree symmetries in networks,'' {\em Physical Review E},
  vol.~74, Sept. 2006.

\bibitem{Sakaguchi1986}
H.~Sakaguchi and Y.~Kuramoto, ``{A Soluble Active Rotator Model Showing Phase
  Transitions via Mutual Entrainment},'' {\em Progress of Theoretical Physics},
  vol.~76, no.~3, pp.~576--581, 1986.

\bibitem{Klickstein2018}
I.~Klickstein and F.~Sorrentino, ``{Generating symmetric graphs},'' {\em
  Chaos}, vol.~28, no.~12, 2018.

\bibitem{Erdos1963}
P.~Erd{\H{o}}s and A.~R{\'{e}}nyi, ``Asymmetric graphs,'' {\em Acta Mathematica
  Academiae Scientiarum Hungaricae}, vol.~14, pp.~295--315, Sept. 1963.

\bibitem{MacArthur2008}
B.~D. MacArthur, R.~J. S{\'{a}}nchez-Garc{\'{i}}a, and J.~W. Anderson,
  ``{Symmetry in complex networks},'' {\em Discrete Applied Mathematics},
  vol.~156, no.~18, pp.~3525--3531, 2008.

\bibitem{Ben2011}
D.~M. Ben, R.~J. S{\'{a}}nchez-Garc{\'{i}}a, and J.~Anderson, ``{Symmetry in
  complex networks},'' {\em Symmetry}, vol.~3, no.~1, pp.~1--15, 2011.

\bibitem{Alon2007}
U.~Alon, ``{Network motifs: Theory and experimental approaches},'' {\em Nature
  Reviews Genetics}, vol.~8, no.~6, pp.~450--461, 2007.

\bibitem{Schaub2016}
M.~T. Schaub, N.~O'Clery, Y.~N. Billeh, J.~C. Delvenne, R.~Lambiotte, and
  M.~Barahona, ``{Graph partitions and cluster synchronization in networks of
  oscillators},'' {\em Chaos}, vol.~26, no.~9, 2016.

\bibitem{McKay2014}
B.~D. McKay and A.~Piperno, ``Practical graph isomorphism, {II},'' {\em Journal
  of Symbolic Computation}, vol.~60, pp.~94--112, Jan. 2014.

\bibitem{saucy}
``saucy 3.0.'' http://vlsicad.eecs.umich.edu/BK/SAUCY/.

\bibitem{gap}
``Gap - groups, algorithms, programming - a system for computational discrete
  algebra.'' https://www.gap-system.org/.

\bibitem{Kuramoto1975}
Y.~Kuramoto, ``{Self-entrainment of a population of coupled non-linear
  oscillators},'' in {\em Lecture Notes in Physics}, vol.~30, (Berl{\'{i}}n,
  Heidelberg), Springer, 1975.

\bibitem{Acebron2005}
J.~A. Acebr{\'{o}}n, L.~L. Bonilla, C.~J. {P{\'{e}}rez Vicente}, F.~Ritort, and
  R.~Spigler, ``{The Kuramoto model: A simple paradigm for synchronization
  phenomena},'' {\em Reviews of Modern Physics}, vol.~77, 2005.

\bibitem{Arenas2008}
A.~Arenas, A.~D{\'{i}}az-Guilera, J.~Kurths, Y.~Moreno, and C.~Zhou,
  ``{Synchronization in complex networks},'' {\em Physics Reports}, vol.~469,
  no.~3, pp.~93--153, 2008.

\bibitem{Nishikawa2016}
T.~Nishikawa and A.~E. Motter, ``{Symmetric States Requiring System
  Asymmetry},'' {\em Physical Review Letters}, vol.~117, no.~11, pp.~1--5,
  2016.

\bibitem{Rosell-Tarrago2020}
G.~Rosell-Tarrag{\'{o}} and A.~D{\'{i}}az-Guilera, ``{Functionability in
  complex networks: Leading nodes for the transition from structural to
  functional networks through remote asynchronization},'' {\em Chaos}, vol.~30,
  no.~1, 2020.

\bibitem{Mitra2006}
N.~J. Mitra, L.~J. Guibas, and M.~Pauly, ``{Partial and approximate symmetry
  detection for 3D geometry},'' {\em ACM Transactions on Graphics}, vol.~25,
  no.~3, pp.~560--568, 2006.

\bibitem{Pakdemirli2004}
M.~Pakdemirli, M.~Y{\"{u}}r{\"{u}}soy, and I.~T. Dolap{\c{c}}i, ``{Comparison
  of Approximate Symmetry Methods for Differential Equations},'' {\em Acta
  Applicandae Mathematicae}, vol.~80, no.~3, pp.~243--271, 2004.

\bibitem{Hall1991}
P.~Hall, S.~J. Sheater, M.~C. Jones, and J.~S. Marron, ``On optimal data-based
  bandwidth selection in kernel density estimation,'' {\em Biometrika},
  vol.~78, no.~2, pp.~263--269, 1991.

\bibitem{Taylor1989}
T.~Charles~C, ``Bootstrap choice of the smoothing parameter in kernel density
  estimation,'' {\em Biometrika}, vol.~76, no.~4, pp.~705--712, 1989.

\bibitem{macaque2018}
K.~Shen, G.~Bezgin, S.~Everling, and A.~R. McIntosh, ``The virtual macaque
  brain: A macaque connectome for large-scale network simulations in
  thevirtualbrain,'' 2018.

\bibitem{Fruchterman1991}
T.~M.~J. Fruchterman and E.~M. Reingold, ``Graph drawing by force-directed
  placement,'' {\em Software: Practice and Experience}, vol.~21,
  pp.~1129--1164, Nov. 1991.

\bibitem{Ketchen1996}
D.~J. Ketchen and C.~L. Shook, ``The application of cluster analysis in
  strategic manegement reserach: an analysis and critique,'' {\em Strategic
  Management Journal}, vol.~17, no.~6, pp.~441--458, 1996.

\bibitem{Rosenblatt1956}
M.~Rosenblatt, ``Remarks on some nonparametric estimates of a density
  function,'' {\em The Annals of Mathematical Statistics}, vol.~27,
  pp.~832--837, Sept. 1956.

\bibitem{Parzen1962}
E.~Parzen, ``On estimation of a probability density function and mode,'' {\em
  The Annals of Mathematical Statistics}, vol.~33, pp.~1065--1076, Sept. 1962.

\bibitem{Turlach1993}
B.~A. Turlach, ``Bandwidth selection in kernel density estimation: A review,''
  in {\em CORE and Institut de Statistique}, 1993.

\bibitem{Botev2010}
Z.~I. Botev, J.~F. Grotowski, and D.~P. Kroese, ``Kernel density estimation via
  diffusion,'' {\em The Annals of Statistics}, vol.~38, pp.~2916--2957, Oct.
  2010.

\bibitem{Terrell1992}
G.~R. Terrell and D.~W. Scott, ``Variable kernel density estimation,'' {\em The
  Annals of Statistics}, vol.~20, no.~3, pp.~1236--1265, 1992.

\end{thebibliography}

\end{document}